\documentclass[aps,prl,twocolumn,superscriptaddress,a4paper]{revtex4-1}
\usepackage[dvipdfmx]{graphicx}

\usepackage{amsmath,amsthm,amssymb}
\usepackage{amsmath,bm}
\usepackage{color,ulem}
\usepackage{ifthen}

\usepackage{multirow,bigdelim}

\newcount \commentout
\newcommand{\1}{\mbox{1}\hspace{-0.25em}\mbox{l}}

\def\tr{\mathrm{tr}}

\newlength{\figwidth}
\newlength{\figlarge}
\begin{document}
\title{
Bulk-edge correspondence of classical diffusion phenomena
}

\author{Tsuneya Yoshida}
\author{Yasuhiro Hatsugai}
\affiliation{Department of Physics, University of Tsukuba, Ibaraki 305-8571, Japan}
\date{\today}
\begin{abstract}
We elucidate that the diffusive systems, which are widely found in nature, can be a new platform of the bulk-edge correspondence, a representative topological phenomenon.
Using a discretized diffusion equation, we demonstrate the emergence of robust edge states protected by the winding number for one- and two-dimensional systems.
These topological edge states can be experimentally accessible by measuring the diffusive dynamics at the edges. 
Furthermore, we discover a novel diffusive phenomenon by numerically simulating the distribution of temperatures for a honeycomb lattice system; the temperature field with wavenumber $\pi$ cannot diffuse to the bulk, which is attributed to the complete localization of the edge state.
\end{abstract}
\maketitle

\textit{
Introduction.--
}
%
In these decades, the notion of topology in condensed matter physics enhances its significance~\cite{Kane_Z2TI_PRL05_1,Kane_Z2TI_PRL05_2,HgTe_Bernevig06,Konig_QSHE2007,Qi_TQFTofTI_PRB08,TI_review_Hasan10,TI_review_Qi10}. 
One of the characteristic topological phenomena is the emergence of robust gapless edge states due to topological properties in the bulk which is known as the bulk-edge correspondence; the chiral edge states emerge~\cite{Halperin_PRB82} corresponding to a finite value of the Chern number in the bulk of the system without symmetry~\cite{Thouless_PRL1982}, which is elucidated in Ref.~\onlinecite{Hatsugai_PRL93}.
The topologically protected edge states are sources of novel phenomena, such as the quantized Hall conductance~\cite{Klitzing_IQHE_PRL80,Thouless_PRL1982}, the emergence of Majorana fermions~\cite{Kitaev_chain_01,Majorana_Ryu_RPL02,Majorana_MourikScience12,Majorana_Rokhinson2012,Majorana_Das2012,Alicia_Majorana_review12,Sato_JPSJ16}, etc.

Remarkably, recent works extended the bulk-edge correspondence to several classical systems which are governed by Maxwell equations, Newton equation, etc.~\cite{Haldane_chiralPHC_PRL08,Raghu_chiralPHC_PRA08,Wang_chiralPHC_Nature09,Ozawa_TopoPhoto_RMP19,ProdanPRL09,Kane_NatPhys13,Kariyado_SR15,Suesstrunk_Mech-class_PNAS16,Chien_Th_Ph_PRB18,Yoshida_SPERs_mech19,Wakao_HOTImech_PRB20,Victor_Topoelecircit_PRL15,Lee_Topoelecircit_CommPhys18,Helbig_ExpSkin_19,Yoshida_MSkinPRR20,Delplace_topoEq_Science17,ActiveMatter_SonePRL19}.
These progresses beyond quantum systems provide universal understanding from the topology and result in invention of new devises (e.g., the topological laser~\cite{Harari_TopoLaser_Science18,Banders_TopoLaser_Science18}) thanks to the robust edge states.
Therefore, further extending the bulk-edge correspondence beyond quantum systems is considered to be significant in term of both the scientific viewpoint and applications.

In this letter, we point out that classical diffusive systems can be a new platform of the bulk-edge correspondence, which highlights topological aspects of the classical diffusive phenomena; the diffusive systems include a wide variety of systems (e.g., thermal diffusion~\cite{Ogi_DiffWavePRL16,LiSciencePT19}, diffusion of impurities in metals~\cite{Peterson_DiffImpMetal_PRB1970}, diffusion of droplets of inks in water, etc.).
To this aim, we discretize the diffusion equation based on Fick's law.
The discretized diffusion equation allows us to discuss the bulk-edge correspondence of diffusion phenomena for the classical systems; the governing equation is expressed in a matrix form that is mathematically equivalent to a tight-binding model of a quantum system.
Our numerical data verify the bulk-edge correspondence for diffusive phenomena in the classical systems.
Furthermore, our numerical simulation of the temperature distribution elucidates a novel diffusive phenomenon for a honeycomb lattice system; 
the temperature field with wavenumber $k_x=\pi$ cannot diffuse to the bulk, which is attributed to the complete localization of the edge state with $k_x=\pi$.

\textit{
Discretizing the diffusion equation.--
}
%
We introduce a discretized diffusion equation [see e.g., Eq.~(\ref{eq: 1D chain fick})] based on Fick's law.

Before addressing the discretization, let us briefly review Fick's law and the diffusion equation of a continuum scalar field $\phi(t,x)$ in one dimension
\begin{eqnarray}
 \label{eq: conn diff}
 \partial_t \phi(t,x) &=&  D \partial^2_x \phi (t,x),
\end{eqnarray}
where $\partial_{t(x)}$ denotes derivative with respect to time $t$ (spatial coordinate $x$).
Here, depending on the system, the scalar field $\phi(t,x)$ corresponds to the field of temperatures, the density of the diffusing material, etc..
Fick's law indicates that the corresponding flux $J$ is given by $J = -D\partial_x \phi(t,x)$, where $D$ is the diffusion coefficient.
By combining this equation and the equation of continuity $\partial_t \phi(t,x)+\partial_x J (t,x)=0$, we obtain the diffusion equation~(\ref{eq: conn diff}).

Now, let us discretize the diffusion equation~(\ref{eq: conn diff}) connecting the diffusion phenomena to tight-binding models of quantum systems.
In order to show the essential idea, we focus on one-dimensional systems.

\begin{figure}[!h]
\begin{minipage}{1\hsize}
\begin{center}
\includegraphics[width=1\hsize,clip]{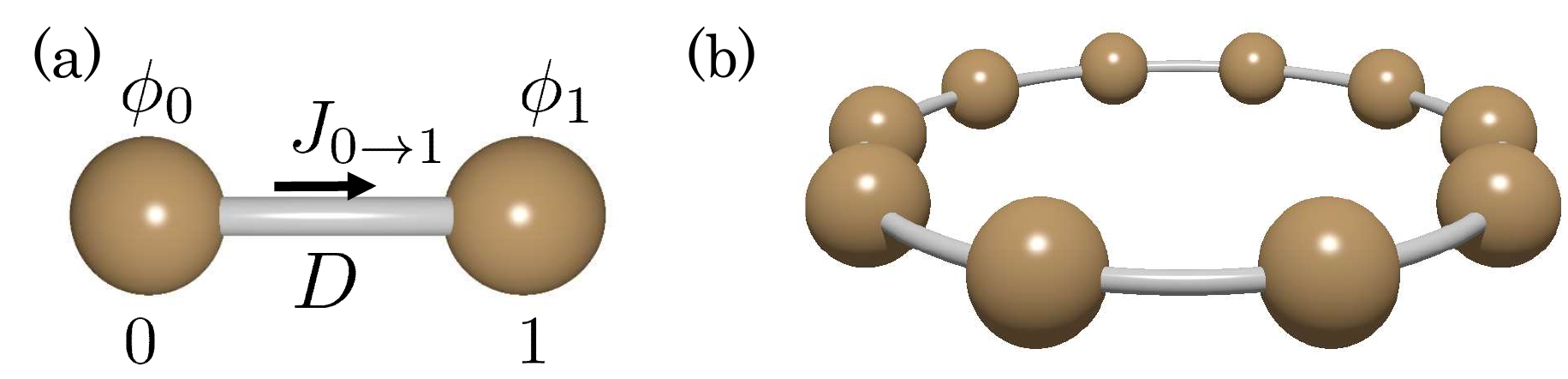}
\end{center}
\end{minipage}
\caption{
(Color Online).
Sketch of the one-dimensional system. 
(a) System composed of two sites coupled with the diffusion coefficient $D$; the flux flowing from site $0$ to site $1$ is written as $\vec{J}_{0\to1}=-D (\phi_0-\phi_1)$, where $\phi$'s denote the discretized field.
(b) One-dimensional chain under the periodic boundary condition for $L_x=10$.
}
\label{fig: model_fick}
\end{figure}

Consider a system composed of two sites where the values of the discretized field $\phi_{0}$ and $\phi_{1}$ are assigned at each site [see Fig.~\ref{fig: model_fick}(a)]; for the heat conduction equation, consider two balls (e.g., macroscopic iron balls) where temperatures are $T_0$ and $T_1$.
Recalling Fick's law, we can write the flux flowing from site $0$ to $1$ with $\phi$'s, $J_{0\to 1} = -D (\phi_0-\phi_1)$. 
Here, we have chosen the distance between the sites as the unit of length.
Thus, the time-evolution of the vector $\vec{\phi}=(\phi_0, \phi_1)^T$ is described by 
\begin{eqnarray}
\partial_t \vec{\phi}(t)
&=& 
-D
\left(
\begin{array}{cc}
1 & -1 \\
-1 & 1
\end{array}
\right)
\vec{\phi}(t).
\end{eqnarray}

Therefore, for a one-dimensional chain composed of $L_x$ sites [see Fig.~\ref{fig: model_fick}(b)], the time-evolution of the vector $\vec{\phi}=(\phi_0, \phi_1,\cdots,\phi_{L_x-1})^T$ is described by
\begin{subequations}
\label{eq: 1D chain fick}
\begin{eqnarray}
\partial_t \vec{\phi}(t)
&=& -\hat{H}\vec{\phi}(t),
\end{eqnarray}
\begin{eqnarray}
\hat{H}
&=& D
\left(
\begin{array}{ccccc}
2      &-1       & 0     & \cdots & -1 \\
-1     & 2       & -1    & \cdots &  0 \\
0      & -1      & 2     & \cdots &  0 \\
\vdots & \vdots  &\vdots & \ddots & \vdots \\
-1     & 0       & 0     & \cdots & 2
\end{array}
\right),
\end{eqnarray}
\end{subequations}
which is a discretized form of the diffusion equation~(\ref{eq: conn diff}).
Here, we have imposed the periodic boundary condition. 
Equation~(\ref{eq: 1D chain fick}) bridges the diffusion phenomena and quantum systems; the matrix $\hat{H}$ corresponds to the Hamiltonian of a one-dimensional tight-binding model.

We note that in the continuum limit, Eq.~(\ref{eq: 1D chain fick}) is reduced to Eq.~(\ref{eq: conn diff}).
To see this, we diagonalize the matrix $\hat{H}$ and focus on the long-wavelength limit.
By applying the Fourier transformation, $\phi_{j_x} = \frac{1}{\sqrt{L_x}} \sum_{k_x} e^{ik_x j_x} \phi_{k_x}$, we obtain the eigenvalues as $\epsilon(k_x)=D(2-2\cos k_x )$ with $k_x=2\pi n_x/L_x$ ($n_x=0,1,\cdots,L_x-1$).
For $k_x\sim 0$, we have $\epsilon(k_x)\sim Dk^2_x$, meaning that the time-evolution is described by Eq.~(\ref{eq: conn diff}) in the long-wavelength limit.
Here we have used the correspondence $k_x \leftrightarrow  -i \partial_x$.

In the above, by discretizing the diffusion equation, we have shown that the diffusive dynamics of classical systems can be described by the tight-binding model of quantum systems [see Eq.~(\ref{eq: 1D chain fick})]. 
This result implies that the diffusive systems serve as a new platform of topological physics beyond quantum systems.

\begin{figure}[!h]
\begin{minipage}{1\hsize}
\begin{center}
\includegraphics[width=1\hsize,clip]{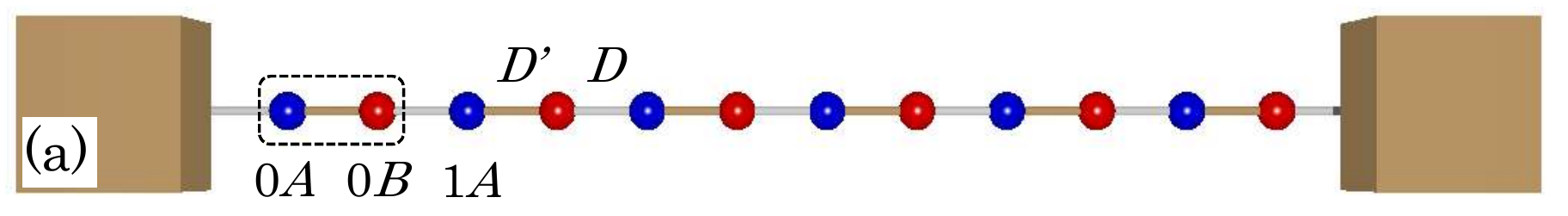}
\end{center}
\end{minipage}
\begin{minipage}{1\hsize}
\begin{center}
\includegraphics[width=1\hsize,clip]{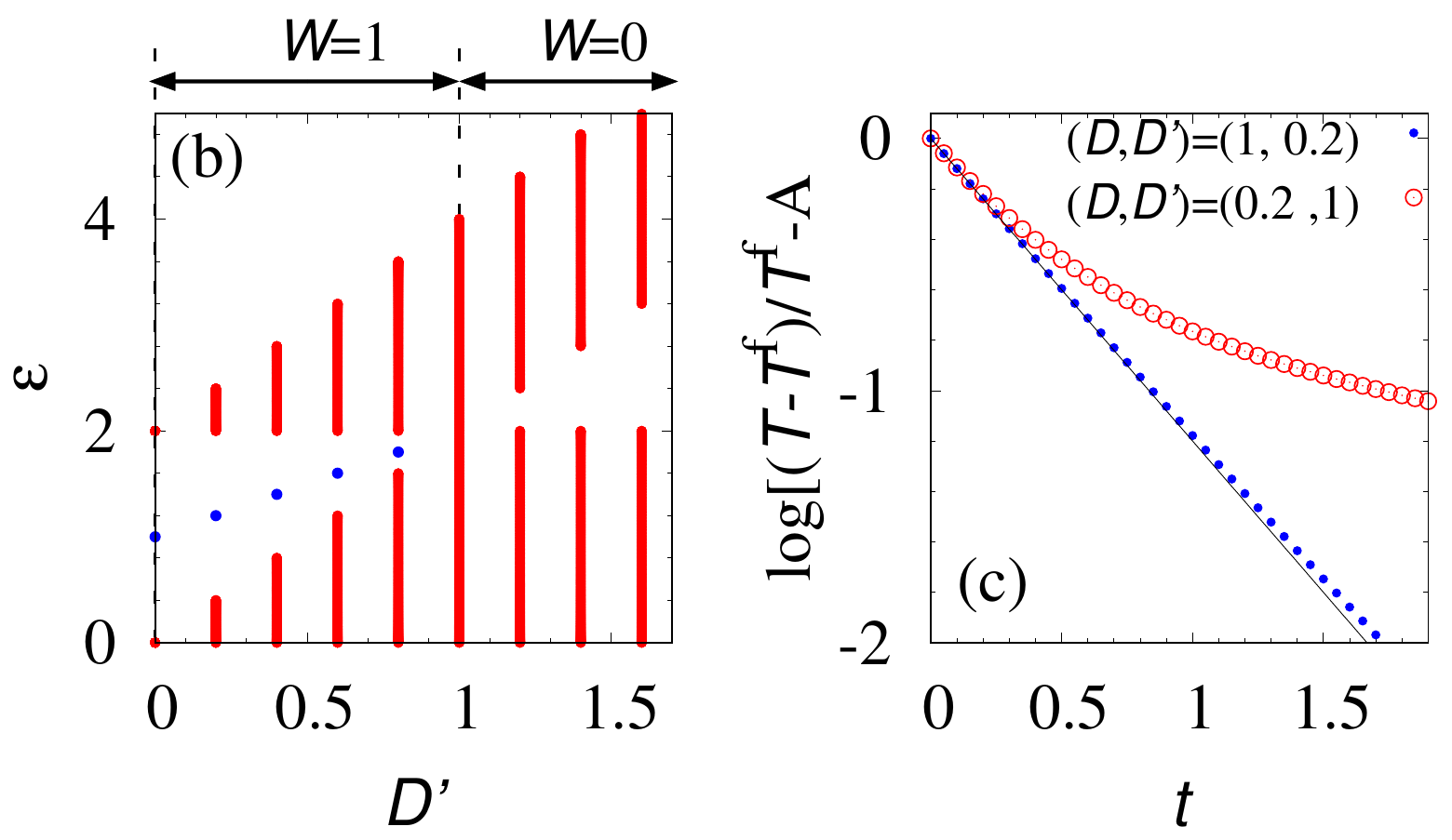}
\end{center}
\end{minipage}
\caption{
(Color Online).
(a): Sketch of the model under the fixed boundary condition for $L_x=6$. 
The sites labeled by $(i_x,\alpha)$ are coupled to the neighboring sites or walls whose coupling strength is denoted by the diffusion coefficient $D'$ (brown) or $D$ (gray). 
We assume that the heat capacity of the wall is sufficiently large.
(b): Spectrum of $\hat{H}_{\mathrm{SSH}}$ for $D=1$ and $L_x=240$. Here, the fixed boundary condition is imposed. For $ 0\leq D'<1$ the system shows the edge states denoted by blue dots because of the bulk topological properties.
(c): The time-evolution of $\vec{T}_{0A}(t)$ in the case for $(D,D')=(1,0.2)$ [$(D,D')=(0.2,1)$] where the system is topologically nontrivial (trivial).
The function $-(D+D')t$ is plotted with a black line.
We have subtracted $A=\log[(T_{0A}(t=0)-T^{ \mathrm{f} })/T^{ \mathrm{f} }]$ for comparison.
The data are obtained with $T^{ \mathrm{f} }=T_{0A}(t=50)$ and $L_x=240$.
We set the initial state as $\vec{T}_{i\alpha}=\delta_{i0}\delta_{\alpha A}$.
}
\label{fig: SSH_spec}
\end{figure}

\textit{
SSH model of the heat conduction equation.--
}
In order to demonstrate that the diffusive dynamics of classical systems indeed show topological phenomena we analyze a one-dimensional system with dimerization [see Fig.~\ref{fig: SSH_spec}(a)] which corresponds to the Su-Schrieffer-Heeger (SSH) model~\cite{SSH_PRL79,SSH_RMP88} of quantum systems.
In the rest of this paper, we discuss the discretized version of the heat conduction equation for the sake of concreteness.

Let us consider the one-dimensional system illustrated in Fig.~\ref{fig: SSH_spec}(a).
The temperature at each site is described by the following vector,
$ 
\vec{T}=
\left(
\begin{array}{ccccc}
T_{0A} & T_{0B} & T_{1A} & \cdots & T_{L_x-1B}
\end{array}
\right)
$. 
Here, the temperature at each site $T_{i_x\alpha}$ ($\alpha=A,B$) is defined as the difference from the temperature of the wall $T_{\mathrm{w}}$.

In a similar way to derive Eq.~(\ref{eq: 1D chain fick}), we obtain the following equation
\begin{eqnarray}
\label{eq: diff SSH}
\partial_t \vec{T}(t) &=& -\hat{H}_{\mathrm{SSH}}\vec{T}(t),
\end{eqnarray}
with $\delta:=D'/D>0$.
For details of the derivation and the specific form of the matrix $\hat{H}_{\mathrm{SSH}}$, see Sec.~\ref{sec: SSH app} of Supplemental Material~\onlinecite{supple}.

Firstly, let us discuss the topological properties in the bulk.
In the momentum space, the matrix $\hat{H}_{\mathrm{SSH}}$ is rewritten as
\begin{eqnarray}
\hat{h}_{\mathrm{SSH}}(k_x)&=& 
D
\left(
\begin{array}{cc}
1+\delta & \delta+e^{i k_x}  \\
\delta+e^{-i k_x} & 1+\delta
\end{array}
\right)_\sigma,
\end{eqnarray}
with $k_x=2\pi n_x /L_x$ and $n_x=0,1,\dots, L_x-1$.
Here, the Pauli matrices act on the sublattice degrees of freedom.
Before analyzing the topological properties, we note that the system shows a gap and preserves the chiral symmetry.
Diagonalizing the matrix, we obtain the spectrum $\epsilon_{\pm}(k_x)=D \left[ (1+\delta) \pm \sqrt{(\delta+\cos k_x)^2+\sin^2 k_x} \right]$.
This result indicates that the spectrum shows a gap for $\delta \neq 1$.
The system also preserves the chiral symmetry; $\hat{h}'_{\mathrm{SSH}}:=\hat{h}_{\mathrm{SSH}}-D(1+\delta)\sigma_0$ satisfies $\sigma_3 \hat{h}'_{\mathrm{SSH}}(k_x) \sigma_3=-\hat{h}'_{\mathrm{SSH}}(k_x)$.
Here, we note that the shift described by the identity matrix $\sigma_0$ does not affect the eigenvalue problem, meaning that topological properties of the eigenvectors are encoded into $\hat{h}'_{\mathrm{SSH}}$.

Because $\hat{h}'_{\mathrm{SSH}}$ shows the gap and preserves the chiral symmetry, it may possesses the topologically nontrivial properties which are characterized by the winding number:
\begin{eqnarray}
W&=& -\int^\pi_{-\pi} \frac{dk_x}{4\pi i} \tr[\sigma_3 \hat{h}'^{-1}_{\mathrm{SSH}}(k_x) \partial_{k_x} \hat{h}'_{\mathrm{SSH}}(k_x)].
\end{eqnarray}
Computing the winding number, we can see that the winding number takes one ($W=1$) for $0 \leq D' < 1$ while it takes zero ($W=0$) for $1 \leq D'$.

For one-dimensional quantum systems with chiral symmetry, the winding number predicts the number of the gapless edge modes localized around the edges, which is typical example of the bulk-edge correspondence.
We show that the bulk-boundary correspondence can be observed in our classical system.
Figure~\ref{fig: SSH_spec}(b) shows the spectrum of $\hat{H}_{\mathrm{SSH}}$ under the fixed boundary condition.
This figure indicates that corresponding the winding number $W=1$ ($W=0$), there exists an edge state (no edge state) localized at each edge, which is represented as a blue dot for each value of $D'$.
Here, the edge state appears at $\epsilon=D+D'$ because of the term proportional to the identity matrix.
We note that the edge states survive even in the presence of the perturbation preserving the chiral symmetry, which supports that the edge states are protected by the topological properties in the bulk.

The above results demonstrate that the diffusive dynamics of classical systems exhibit the bulk-edge correspondence which is a unique topological phenomenon.

\textit{
How to experimentally access the edge states.--
}
So far, we have shown that the edge states emerge at $\epsilon=D+D'$ because of the topological properties in the bulk. 
In the following, let us discuss how to experimentally access the edge states.

One possibility is to observe the time-evolution of the temperature at the edge which is consider to decay exponentially $T_{0A}\sim e^{-(D+D')t}$. 
In Fig.~\ref{fig: SSH_spec}(c), the time-evolution of the temperature at edge $(i_x,\alpha)=(0,A)$ is plotted.
The temperature $T_{0A}$ shows exponential decay for $t\lesssim 2 \tau$ with the half-life $\tau=1/(D+D')=0.83$ for $(D,D')=(1,0.2)$ due to the edge state, while it deviates from the line of the exponential decay around $t=0.5$ which is shorter than the half-life for $(D,D')=(0.2,1)$.
The above behaviors due to the emergence of the edge states can be observed even in the presence of the disorder.
Therefore, we conclude that observing the time-evolution allows us to experimentally access the edge states induced by the bulk topological properties.
We note that the time-evolution of the temperature at each site has been measured in Ref.~\onlinecite{Ogi_DiffWavePRL16} for continuous systems~\cite{estimate_ftnt}.

We also consider that at least in principle, the eigenvectors and eigenvalues of the matrix $\hat{H}_{\mathrm{SSH}}$ can be extracted from the experimental data in the following procedure.
(i) Prepare a set of initial conditions $\vec{T}^{(\mathrm{i})}(t=0)_{l}$ ($l=0,\cdots,L_x-1$) which are linear independent each other; for instance, such initial conditions can be prepared by heating at a site.
(ii) Observe the temperature $\vec{T}^{(\mathrm{f})}_{l}$ at time $t_0$ for each case of initial condition. 
Here, these two sets of experimental data satisfy 
\begin{eqnarray}
\label{eq: Hssh_exp}
\hat{T}(t_0) &=& e^{-H_{\mathrm{SSH}} t_0} \hat{T}(0),
\end{eqnarray}
with $\hat{T}(t_0)=(\vec{T}^{(\mathrm{f})}_{0},\vec{T}^{(\mathrm{f})}_{1}, \cdots, \vec{T}^{(\mathrm{f})}_{L_x-1})$ and $\hat{T}(0)=(\vec{T}^{(\mathrm{i})}_{0},\vec{T}^{(\mathrm{i})}_{1}, \cdots, \vec{T}^{(\mathrm{i})}_{L_x-1})$.
(iii) Diagonalizing $\hat{T}(t_0)[\hat{T}(0)]^{-1}$, which is identical to $e^{-\hat{H}_{\mathrm{SSH}} t_0}$, we obtain the eigenvalues and eigenstates of $\hat{H}_{\mathrm{SSH}}$.

\begin{figure}[!h]
\begin{minipage}{1\hsize}
\begin{center}
\includegraphics[width=1\hsize,clip]{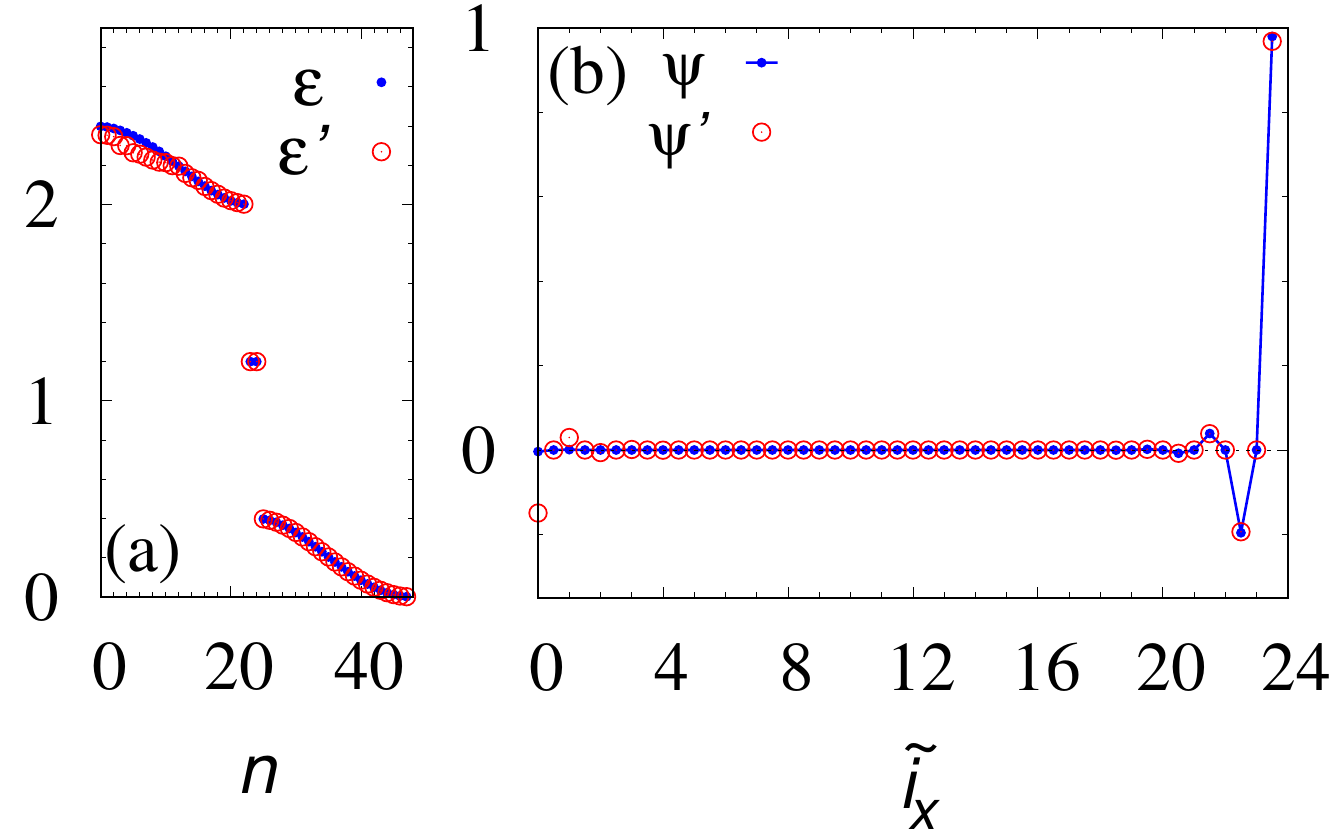}
\end{center}
\end{minipage}
\caption{
(Color Online).
(a) [(b)] The eigenvalues (the edge state) obtained from $\hat{H}_{\mathrm{SSH}}$ and $\hat{T}(t_0)[\hat{T}(0)]^{-1}$ for $D=1$, $D'=0.2$, $L_x=24$, and $t_0=17$.
The eigenvalues of the matrix $\hat{T}(t_0)[\hat{T}(0)]^{-1}$ are defined as $e^{-\epsilon'_n t_0}$.
The set of labels $(i_x,\alpha)$ is represented as $\tilde{i}_x$ as follows: $\tilde{i}_x$ takes $i_x$ [$i_x+0.5$] for $(i_x,A)$ [$(i_x,B)$].
The deviation for $ 2.5 \lesssim \epsilon \lesssim 3$ observed in panel (a) is due to the fact that the matrix element of $\hat{T}(t_0)$ become small for large $t_0$.
}
\label{fig: TheExp}
\end{figure}

Figure~\ref{fig: TheExp}(a) shows eigenvalues of $\hat{T}(t_0)[\hat{T}(0)]^{-1}$. 
The eigenvalues are obtained with the initial condition $[\vec{T}^{(\mathrm{i})}_l]_{i_x\alpha}=T_{l\alpha}\delta_{li_x}\delta_{{\alpha_l}\alpha}$ with $\alpha_l=A,B$ and $T_{l\alpha}$ taking a random value~\cite{T_is_au_ftnt} between $0.5$ and $1$.
The eigenvalues $e^{-\epsilon' t_0}$ almost reproduce the ones of $\hat{H}_{\mathrm{SSH}}$.
We note that the deviation for $ 2.5 \lesssim \epsilon \lesssim 3$ is due to the rounding error; the matrix elements of $\hat{T}(t_0)_{ji}$ exponentially decay.
Figure~\ref{fig: TheExp}(b) shows the eigenvector of $\hat{T}(t_0)[\hat{T}(0)]^{-1}$ which corresponds to the edge mode.
The eigenvector also is in nice agreement with the edge state of $\hat{H}_{\mathrm{SSH}}$.

\begin{figure}[!h]
\begin{minipage}{1\hsize}
\begin{center}
\includegraphics[width=0.8\hsize,clip]{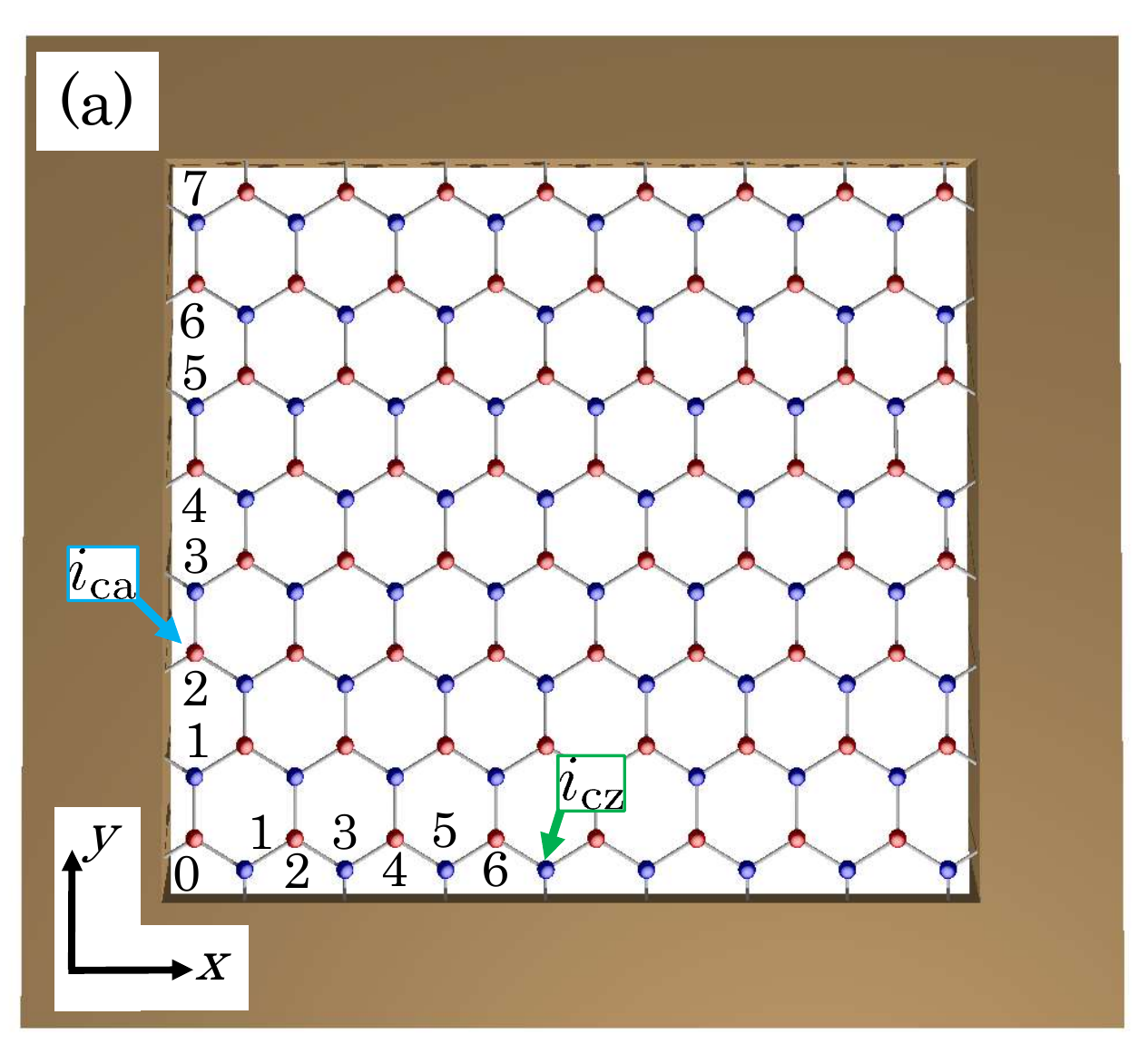}
\end{center}
\end{minipage}
\begin{minipage}{1\hsize}
\begin{center}
\includegraphics[width=1\hsize,clip]{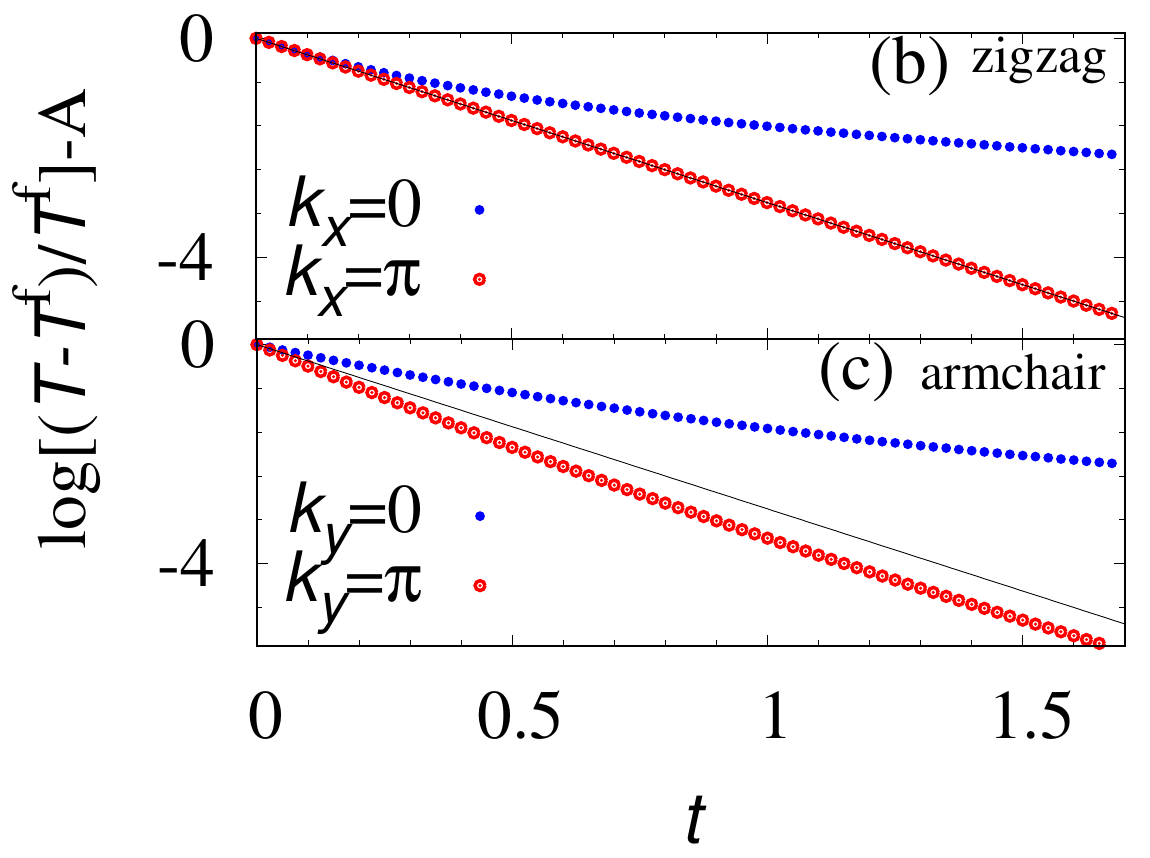}
\end{center}
\end{minipage}
\caption{
(Color Online). 
(a) Sketch of the honeycomb lattice for $L_x=16$ and $L_y=8$.
Here, the fixed boundary condition is imposed both for the $x$- and $y$-directions. 
The numbers along the $x$- ($y$-) direction represent $i_x=0,1,2,\cdots,15$ ($i_y=0,1,2,\cdots,7$).
(b) [(c)] Time-evolution of temperature $T_{i_{\mathrm{cz}}}$ [$T_{i_{\mathrm{ca}}}$] along a zigzag edge [an armchair edge] for $L_x=L_y=40$ and $D=1$.
The subscript $i_{\mathrm{cz}}=(L_x/2-1,0)$ [$i_{\mathrm{ca}}=(0,L_y/2-2)$] specifies the site on the zigzag [armchair] edge. For $L_x=16$ and $L_y=8$, the site specified by  $i_{\mathrm{cz}}=(7,0)$ [$i_{\mathrm{ca}}=(0,2)$] is denoted by the green (blue) arrow in panel (a).
The function $-3Dt$ is plotted with a black line.
The temperature $T^{\mathrm{f}}$ is set to $T_{i_{\mathrm{cz}}}(t=50)$ [$T_{i_{\mathrm{ca}}}(t=50)$] for the zigzag (armchair) edge.
We have subtracted $A=\log[(T_{i_{\mathrm{cz}}(i_{\mathrm{ca}})}(t=0)-T^{\mathrm{f}})/T^{\mathrm{f}}]$ for comparison.
Data shown in panel (b) [(c)] are obtained with the initial condition whose wave number is $k_x=0$ or $\pi$ [$k_y=0$ or $\pi$].
Here, the boundary conditions are imposed as follows: for panel (b) [(c)] the periodic and fixed [fixed and periodic] boundary conditions are imposed for the $x$- and $y$-directions, respectively.
}
\label{fig: honeycomb}
\end{figure}

\textit{
Honeycomb lattice system.--
}
Topological phenomena of the diffusive dynamics can also be found for two-dimensional systems.
To show this, we analyze a honeycomb lattice system illustrated in Fig.~\ref{fig: honeycomb}(a) where the fixed boundary condition is imposed both for the $x$- and $y$-directions. 
We have supposed that the sites are coupled with the diffusion coefficient $D$.
The dynamics of the temperature at each site $\vec{T}$ is described by $\partial_t \vec{T} =-\hat{H}_{\mathrm{honey}} \vec{T}$.
As is the case of the SSH model, $\hat{H}_{\mathrm{honey}}$ corresponds to the honeycomb lattice of the tight-binding model; $\hat{H}'_{\mathrm{honey}}:=\hat{H}_{\mathrm{honey}}-3D\1$ preserves the chiral symmetry.

Under the periodic (fixed) boundary condition for the $x$- ($y$-) direction, the system can be regarded as a set of one-dimensional system aligned along the momentum space $ -\pi \leq k_x < \pi$.
Noting that the one-dimensional system specified by $k_x$ preserves the chiral symmetry, we can compute the winding number; the winding number takes one ($W=1$) for $2\pi/3< |k_x| < \pi $, while it takes zero ($W=0$) for $0 \leq  |k_x| < 2\pi/3$.
Correspondingly, only for $2\pi/3<|k_x|<\pi$, the edge state appears~\cite{Fujita_FujitaState_JPSJ96,Majorana_Ryu_RPL02}.
We note that along the armchair edge, no edge states can be observed. For more details of the spectrum, see Sec.~\ref{sec: honey app} of Supplemental Material~\onlinecite{supple}.

The presence or absence of the edge state can affect the diffusive dynamics. 
Figure~\ref{fig: honeycomb}(b)[(c)] shows the time-evolution at site $i_{\mathrm{cz}}$ ($i_{\mathrm{az}}$).
Figure~\ref{fig: honeycomb}(b) shows the dynamics obtained for the two cases of the initial condition spatially modulating either $k_x=0$ or $k_x=\pi$.
The data of $k_x=\pi$ are obtained by subtracting data obtained with the initial condition $2\vec{T}_{\mathrm{iz1}}$ from the ones obtained with $\vec{T}_{\mathrm{iz2}}$ 
(For specific form of $\vec{T}_{\mathrm{iz1}}$ and $\vec{T}_{\mathrm{iz2}}$, see Sec.~\ref{sec: honey app} of Supplemental Material~\onlinecite{supple}).
Figure~\ref{fig: honeycomb}(b) indicates that at the zigzag edge, the temperature field with $k_x=\pi$ exponentially decays $T_{\mathrm{icz}}\sim e^{-3Dt}$ while the data of the temperature field with $k_x=0$ deviates from $e^{-3Dt}$.
The above time-evolution is consistent with the presence of the edge state for $2\pi/3 < |k_x| <\pi$ whose eigenvalue is $3D$. 
We note that the time-evolution of the armchair edge deviates from $e^{-3Dt}$ for both cases of initial conditions.
\begin{figure}[!h]
\begin{minipage}{1\hsize}
\begin{center}
\includegraphics[width=1\hsize,clip]{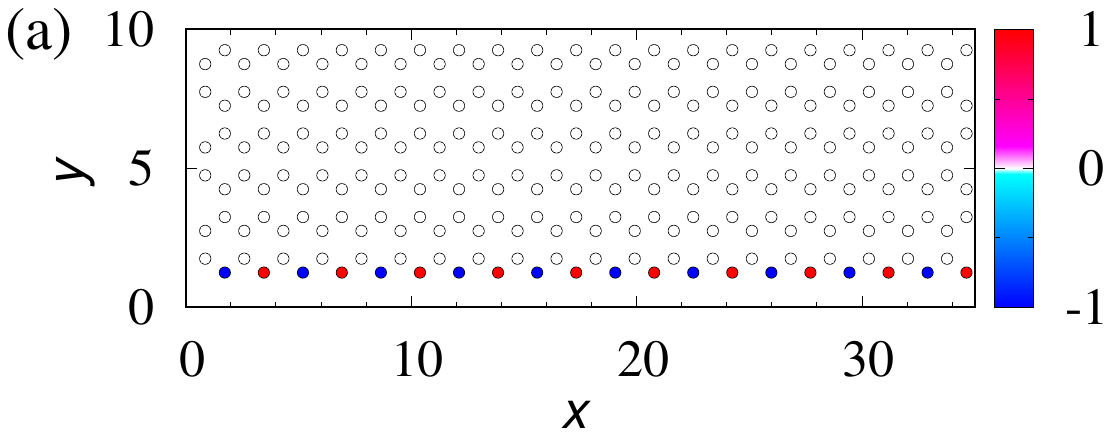}
\end{center}
\end{minipage}
\begin{minipage}{1\hsize}
\begin{center}
\includegraphics[width=1\hsize,clip]{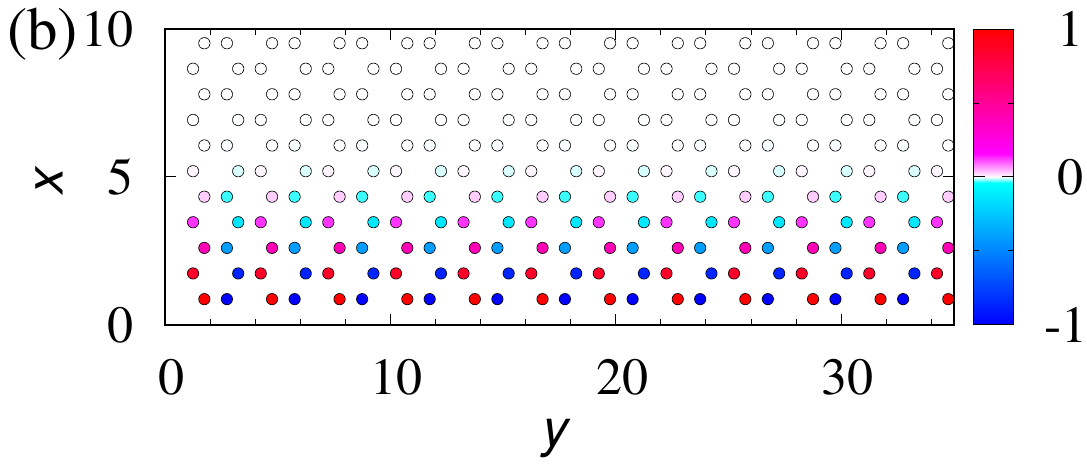}
\end{center}
\end{minipage}
\caption{
(Color Online).
(a) [(b)] Color plot of $\vec{T}(t=1)/T_0$ for a zigzag (an armchair) edge.
The initial condition is chosen as $\vec{T}_{\mathrm{iz3}}$ ($\vec{T}_{\mathrm{ia3}}$) for data of zigzag (armchair) edges, which allows us to observe the mode with $k_x=\pi$ ($k_y=\pi$) for the zigzag (armchair) edge. 
For more details of the initial condition, see Fig.~\ref{fig: honey_ini} and Sec.~\ref{sec: honey app} of Supplemental Material~\onlinecite{supple}.
The data shown in panel (a) [(b)] are obtained under the periodic and fixed (fixed and periodic) boundary conditions along the $x$- and $y$-directions.
Here, we have taken $T_0=0.0497$ ($0.0292$) for data of zigzag (armchair) edges.
The data are obtained for $L_x=L_y=40$ and $D=1$.
}
\label{fig: honeycomb_cmap}
\end{figure}

Furthermore, the edge state at $k_x=\pi$ results in counter intuitive dynamics; for the zigzag edge, the initial state with $k_x=\pi$ cannot diffuse to the bulk [see Fig.~\ref{fig: honeycomb_cmap}(a)] while for the armchair edge, the initial state diffuses to the bulk [see Fig.~\ref{fig: honeycomb_cmap}(b)]. This intriguing behavior is due to the complete localization of the edge state with $k_x=\pi$.
The above counterintuitive behavior is due to the complete localization of the state around the zigzag edge.

\textit{
Summary.--
}
In this letter, we have elucidated the topological aspect of the diffusive dynamics, providing a new platform of the bulk-edge correspondence.

Specifically, based on Fick's law, we have introduced the discretized form of the diffusion equation, bridging the diffusive dynamics of classical systems and a tight-binding model discussed for quantum systems. 
The correspondence between the classical and quantum systems allows us to discuss the topological phenomena (e.g., the bulk-edge correspondence) for the diffusive dynamics of classical systems;
we have numerically elucidated that topological properties characterized by the winding number in the bulk induces the edge states for the one-dimensional system and the honeycomb lattice system.
Furthermore, our numerical simulation has revealed a novel diffusive phenomenon for the honeycomb lattice system; at zigzag edges, the temperature field with spatial modulation $k_x=\pi$ cannot diffuse to the bulk.

Our results provide topological insights into diffusive phenomena, indicating the potential existence of diffusive phenomena analog of topological insulators for other symmetry classes and higher-order insulators.
Their realization is left as future works to be addressed.

\textit{
Acknowledgements.--
}

This work is supported by JSPS Grant-in-Aid for Scientific Research on Innovative Areas ``Discrete Geometric Analysis for Materials Design": Grants No.~JP20H04627.
This work is also supported by JSPS KAKENHI Grants No.~JP17H06138, and No.~JP19K21032.


\begin{thebibliography}{46}%
\makeatletter
\providecommand \@ifxundefined [1]{%
 \@ifx{#1\undefined}
}%
\providecommand \@ifnum [1]{%
 \ifnum #1\expandafter \@firstoftwo
 \else \expandafter \@secondoftwo
 \fi
}%
\providecommand \@ifx [1]{%
 \ifx #1\expandafter \@firstoftwo
 \else \expandafter \@secondoftwo
 \fi
}%
\providecommand \natexlab [1]{#1}%
\providecommand \enquote  [1]{``#1''}%
\providecommand \bibnamefont  [1]{#1}%
\providecommand \bibfnamefont [1]{#1}%
\providecommand \citenamefont [1]{#1}%
\providecommand \href@noop [0]{\@secondoftwo}%
\providecommand \href [0]{\begingroup \@sanitize@url \@href}%
\providecommand \@href[1]{\@@startlink{#1}\@@href}%
\providecommand \@@href[1]{\endgroup#1\@@endlink}%
\providecommand \@sanitize@url [0]{\catcode `\\12\catcode `\$12\catcode
  `\&12\catcode `\#12\catcode `\^12\catcode `\_12\catcode `\%12\relax}%
\providecommand \@@startlink[1]{}%
\providecommand \@@endlink[0]{}%
\providecommand \url  [0]{\begingroup\@sanitize@url \@url }%
\providecommand \@url [1]{\endgroup\@href {#1}{\urlprefix }}%
\providecommand \urlprefix  [0]{URL }%
\providecommand \Eprint [0]{\href }%
\providecommand \doibase [0]{http://dx.doi.org/}%
\providecommand \selectlanguage [0]{\@gobble}%
\providecommand \bibinfo  [0]{\@secondoftwo}%
\providecommand \bibfield  [0]{\@secondoftwo}%
\providecommand \translation [1]{[#1]}%
\providecommand \BibitemOpen [0]{}%
\providecommand \bibitemStop [0]{}%
\providecommand \bibitemNoStop [0]{.\EOS\space}%
\providecommand \EOS [0]{\spacefactor3000\relax}%
\providecommand \BibitemShut  [1]{\csname bibitem#1\endcsname}%
\let\auto@bib@innerbib\@empty
\bibitem [{\citenamefont {Kane}\ and\ \citenamefont
  {Mele}(2005{\natexlab{a}})}]{Kane_Z2TI_PRL05_1}%
  \BibitemOpen
  \bibfield  {author} {\bibinfo {author} {\bibfnamefont {C.~L.}\ \bibnamefont
  {Kane}}\ and\ \bibinfo {author} {\bibfnamefont {E.~J.}\ \bibnamefont
  {Mele}},\ }\href {\doibase 10.1103/PhysRevLett.95.146802} {\bibfield
  {journal} {\bibinfo  {journal} {Phys. Rev. Lett.}\ }\textbf {\bibinfo
  {volume} {95}},\ \bibinfo {pages} {146802} (\bibinfo {year}
  {2005}{\natexlab{a}})}\BibitemShut {NoStop}%
\bibitem [{\citenamefont {Kane}\ and\ \citenamefont
  {Mele}(2005{\natexlab{b}})}]{Kane_Z2TI_PRL05_2}%
  \BibitemOpen
  \bibfield  {author} {\bibinfo {author} {\bibfnamefont {C.~L.}\ \bibnamefont
  {Kane}}\ and\ \bibinfo {author} {\bibfnamefont {E.~J.}\ \bibnamefont
  {Mele}},\ }\href {\doibase 10.1103/PhysRevLett.95.226801} {\bibfield
  {journal} {\bibinfo  {journal} {Phys. Rev. Lett.}\ }\textbf {\bibinfo
  {volume} {95}},\ \bibinfo {pages} {226801} (\bibinfo {year}
  {2005}{\natexlab{b}})}\BibitemShut {NoStop}%
\bibitem [{\citenamefont {Bernevig}\ \emph {et~al.}(2006)\citenamefont
  {Bernevig}, \citenamefont {Hughes},\ and\ \citenamefont
  {Zhang}}]{HgTe_Bernevig06}%
  \BibitemOpen
  \bibfield  {author} {\bibinfo {author} {\bibfnamefont {B.~A.}\ \bibnamefont
  {Bernevig}}, \bibinfo {author} {\bibfnamefont {T.~L.}\ \bibnamefont
  {Hughes}}, \ and\ \bibinfo {author} {\bibfnamefont {S.-C.}\ \bibnamefont
  {Zhang}},\ }\href {\doibase 10.1126/science.1133734} {\bibfield  {journal}
  {\bibinfo  {journal} {Science}\ }\textbf {\bibinfo {volume} {314}},\ \bibinfo
  {pages} {1757} (\bibinfo {year} {2006})}\BibitemShut {NoStop}%
\bibitem [{\citenamefont {K\"onig}\ \emph {et~al.}(2007)\citenamefont
  {K\"onig}, \citenamefont {Wiedmann}, \citenamefont {Br\"une}, \citenamefont
  {Roth}, \citenamefont {Buhmann}, \citenamefont {Molenkamp}, \citenamefont
  {Qi},\ and\ \citenamefont {Zhang}}]{Konig_QSHE2007}%
  \BibitemOpen
  \bibfield  {author} {\bibinfo {author} {\bibfnamefont {M.}~\bibnamefont
  {K\"onig}}, \bibinfo {author} {\bibfnamefont {S.}~\bibnamefont {Wiedmann}},
  \bibinfo {author} {\bibfnamefont {C.}~\bibnamefont {Br\"une}}, \bibinfo
  {author} {\bibfnamefont {A.}~\bibnamefont {Roth}}, \bibinfo {author}
  {\bibfnamefont {H.}~\bibnamefont {Buhmann}}, \bibinfo {author} {\bibfnamefont
  {L.~W.}\ \bibnamefont {Molenkamp}}, \bibinfo {author} {\bibfnamefont {X.-L.}\
  \bibnamefont {Qi}}, \ and\ \bibinfo {author} {\bibfnamefont {S.-C.}\
  \bibnamefont {Zhang}},\ }\href {\doibase 10.1126/science.1148047} {\bibfield
  {journal} {\bibinfo  {journal} {Science}\ }\textbf {\bibinfo {volume}
  {318}},\ \bibinfo {pages} {766} (\bibinfo {year} {2007})}\BibitemShut
  {NoStop}%
\bibitem [{\citenamefont {Qi}\ \emph {et~al.}(2008)\citenamefont {Qi},
  \citenamefont {Hughes},\ and\ \citenamefont {Zhang}}]{Qi_TQFTofTI_PRB08}%
  \BibitemOpen
  \bibfield  {author} {\bibinfo {author} {\bibfnamefont {X.-L.}\ \bibnamefont
  {Qi}}, \bibinfo {author} {\bibfnamefont {T.~L.}\ \bibnamefont {Hughes}}, \
  and\ \bibinfo {author} {\bibfnamefont {S.-C.}\ \bibnamefont {Zhang}},\ }\href
  {\doibase 10.1103/PhysRevB.78.195424} {\bibfield  {journal} {\bibinfo
  {journal} {Phys. Rev. B}\ }\textbf {\bibinfo {volume} {78}},\ \bibinfo
  {pages} {195424} (\bibinfo {year} {2008})}\BibitemShut {NoStop}%
\bibitem [{\citenamefont {Hasan}\ and\ \citenamefont
  {Kane}(2010)}]{TI_review_Hasan10}%
  \BibitemOpen
  \bibfield  {author} {\bibinfo {author} {\bibfnamefont {M.~Z.}\ \bibnamefont
  {Hasan}}\ and\ \bibinfo {author} {\bibfnamefont {C.~L.}\ \bibnamefont
  {Kane}},\ }\href {\doibase 10.1103/RevModPhys.82.3045} {\bibfield  {journal}
  {\bibinfo  {journal} {Rev. Mod. Phys.}\ }\textbf {\bibinfo {volume} {82}},\
  \bibinfo {pages} {3045} (\bibinfo {year} {2010})}\BibitemShut {NoStop}%
\bibitem [{\citenamefont {Qi}\ and\ \citenamefont
  {Zhang}(2011)}]{TI_review_Qi10}%
  \BibitemOpen
  \bibfield  {author} {\bibinfo {author} {\bibfnamefont {X.-L.}\ \bibnamefont
  {Qi}}\ and\ \bibinfo {author} {\bibfnamefont {S.-C.}\ \bibnamefont {Zhang}},\
  }\href {\doibase 10.1103/RevModPhys.83.1057} {\bibfield  {journal} {\bibinfo
  {journal} {Rev. Mod. Phys.}\ }\textbf {\bibinfo {volume} {83}},\ \bibinfo
  {pages} {1057} (\bibinfo {year} {2011})}\BibitemShut {NoStop}%
\bibitem [{\citenamefont {Halperin}(1982)}]{Halperin_PRB82}%
  \BibitemOpen
  \bibfield  {author} {\bibinfo {author} {\bibfnamefont {B.~I.}\ \bibnamefont
  {Halperin}},\ }\href {\doibase 10.1103/PhysRevB.25.2185} {\bibfield
  {journal} {\bibinfo  {journal} {Phys. Rev. B}\ }\textbf {\bibinfo {volume}
  {25}},\ \bibinfo {pages} {2185} (\bibinfo {year} {1982})}\BibitemShut
  {NoStop}%
\bibitem [{\citenamefont {Thouless}\ \emph {et~al.}(1982)\citenamefont
  {Thouless}, \citenamefont {Kohmoto}, \citenamefont {Nightingale},\ and\
  \citenamefont {den Nijs}}]{Thouless_PRL1982}%
  \BibitemOpen
  \bibfield  {author} {\bibinfo {author} {\bibfnamefont {D.~J.}\ \bibnamefont
  {Thouless}}, \bibinfo {author} {\bibfnamefont {M.}~\bibnamefont {Kohmoto}},
  \bibinfo {author} {\bibfnamefont {M.~P.}\ \bibnamefont {Nightingale}}, \ and\
  \bibinfo {author} {\bibfnamefont {M.}~\bibnamefont {den Nijs}},\ }\href
  {\doibase 10.1103/PhysRevLett.49.405} {\bibfield  {journal} {\bibinfo
  {journal} {Phys. Rev. Lett.}\ }\textbf {\bibinfo {volume} {49}},\ \bibinfo
  {pages} {405} (\bibinfo {year} {1982})}\BibitemShut {NoStop}%
\bibitem [{\citenamefont {Hatsugai}(1993)}]{Hatsugai_PRL93}%
  \BibitemOpen
  \bibfield  {author} {\bibinfo {author} {\bibfnamefont {Y.}~\bibnamefont
  {Hatsugai}},\ }\href {\doibase 10.1103/PhysRevLett.71.3697} {\bibfield
  {journal} {\bibinfo  {journal} {Phys. Rev. Lett.}\ }\textbf {\bibinfo
  {volume} {71}},\ \bibinfo {pages} {3697} (\bibinfo {year}
  {1993})}\BibitemShut {NoStop}%
\bibitem [{\citenamefont {Klitzing}\ \emph {et~al.}(1980)\citenamefont
  {Klitzing}, \citenamefont {Dorda},\ and\ \citenamefont
  {Pepper}}]{Klitzing_IQHE_PRL80}%
  \BibitemOpen
  \bibfield  {author} {\bibinfo {author} {\bibfnamefont {K.~v.}\ \bibnamefont
  {Klitzing}}, \bibinfo {author} {\bibfnamefont {G.}~\bibnamefont {Dorda}}, \
  and\ \bibinfo {author} {\bibfnamefont {M.}~\bibnamefont {Pepper}},\ }\href
  {\doibase 10.1103/PhysRevLett.45.494} {\bibfield  {journal} {\bibinfo
  {journal} {Phys. Rev. Lett.}\ }\textbf {\bibinfo {volume} {45}},\ \bibinfo
  {pages} {494} (\bibinfo {year} {1980})}\BibitemShut {NoStop}%
\bibitem [{\citenamefont {Kitaev}(2001)}]{Kitaev_chain_01}%
  \BibitemOpen
  \bibfield  {author} {\bibinfo {author} {\bibfnamefont {A.~Y.}\ \bibnamefont
  {Kitaev}},\ }\href {http://stacks.iop.org/1063-7869/44/i=10S/a=S29}
  {\bibfield  {journal} {\bibinfo  {journal} {Physics-Uspekhi}\ }\textbf
  {\bibinfo {volume} {44}},\ \bibinfo {pages} {131} (\bibinfo {year}
  {2001})}\BibitemShut {NoStop}%
\bibitem [{\citenamefont {Ryu}\ and\ \citenamefont
  {Hatsugai}(2002)}]{Majorana_Ryu_RPL02}%
  \BibitemOpen
  \bibfield  {author} {\bibinfo {author} {\bibfnamefont {S.}~\bibnamefont
  {Ryu}}\ and\ \bibinfo {author} {\bibfnamefont {Y.}~\bibnamefont {Hatsugai}},\
  }\href {\doibase 10.1103/PhysRevLett.89.077002} {\bibfield  {journal}
  {\bibinfo  {journal} {Phys. Rev. Lett.}\ }\textbf {\bibinfo {volume} {89}},\
  \bibinfo {pages} {077002} (\bibinfo {year} {2002})}\BibitemShut {NoStop}%
\bibitem [{\citenamefont {Mourik}\ \emph {et~al.}(2012)\citenamefont {Mourik},
  \citenamefont {Zuo}, \citenamefont {Frolov}, \citenamefont {Plissard},
  \citenamefont {Bakkers},\ and\ \citenamefont
  {Kouwenhoven}}]{Majorana_MourikScience12}%
  \BibitemOpen
  \bibfield  {author} {\bibinfo {author} {\bibfnamefont {V.}~\bibnamefont
  {Mourik}}, \bibinfo {author} {\bibfnamefont {K.}~\bibnamefont {Zuo}},
  \bibinfo {author} {\bibfnamefont {S.~M.}\ \bibnamefont {Frolov}}, \bibinfo
  {author} {\bibfnamefont {S.~R.}\ \bibnamefont {Plissard}}, \bibinfo {author}
  {\bibfnamefont {E.~P. A.~M.}\ \bibnamefont {Bakkers}}, \ and\ \bibinfo
  {author} {\bibfnamefont {L.~P.}\ \bibnamefont {Kouwenhoven}},\ }\href
  {\doibase 10.1126/science.1222360} {\bibfield  {journal} {\bibinfo  {journal}
  {Science}\ }\textbf {\bibinfo {volume} {336}},\ \bibinfo {pages} {1003}
  (\bibinfo {year} {2012})}\BibitemShut {NoStop}%
\bibitem [{\citenamefont {Rokhinson}\ \emph {et~al.}(2012)\citenamefont
  {Rokhinson}, \citenamefont {Liu},\ and\ \citenamefont
  {Furdyna}}]{Majorana_Rokhinson2012}%
  \BibitemOpen
  \bibfield  {author} {\bibinfo {author} {\bibfnamefont {L.~P.}\ \bibnamefont
  {Rokhinson}}, \bibinfo {author} {\bibfnamefont {X.}~\bibnamefont {Liu}}, \
  and\ \bibinfo {author} {\bibfnamefont {J.~K.}\ \bibnamefont {Furdyna}},\
  }\href@noop {} {\bibfield  {journal} {\bibinfo  {journal} {Nature Physics}\
  }\textbf {\bibinfo {volume} {8}},\ \bibinfo {pages} {795} (\bibinfo {year}
  {2012})}\BibitemShut {NoStop}%
\bibitem [{\citenamefont {Das}\ \emph {et~al.}(2012)\citenamefont {Das},
  \citenamefont {Ronen}, \citenamefont {Most}, \citenamefont {Oreg},
  \citenamefont {Heiblum},\ and\ \citenamefont {Shtrikman}}]{Majorana_Das2012}%
  \BibitemOpen
  \bibfield  {author} {\bibinfo {author} {\bibfnamefont {A.}~\bibnamefont
  {Das}}, \bibinfo {author} {\bibfnamefont {Y.}~\bibnamefont {Ronen}}, \bibinfo
  {author} {\bibfnamefont {Y.}~\bibnamefont {Most}}, \bibinfo {author}
  {\bibfnamefont {Y.}~\bibnamefont {Oreg}}, \bibinfo {author} {\bibfnamefont
  {M.}~\bibnamefont {Heiblum}}, \ and\ \bibinfo {author} {\bibfnamefont
  {H.}~\bibnamefont {Shtrikman}},\ }\href@noop {} {\bibfield  {journal}
  {\bibinfo  {journal} {Nature Physics}\ }\textbf {\bibinfo {volume} {8}},\
  \bibinfo {pages} {887} (\bibinfo {year} {2012})}\BibitemShut {NoStop}%
\bibitem [{\citenamefont {Alicea}(2012)}]{Alicia_Majorana_review12}%
  \BibitemOpen
  \bibfield  {author} {\bibinfo {author} {\bibfnamefont {J.}~\bibnamefont
  {Alicea}},\ }\href {http://stacks.iop.org/0034-4885/75/i=7/a=076501}
  {\bibfield  {journal} {\bibinfo  {journal} {Reports on Progress in Physics}\
  }\textbf {\bibinfo {volume} {75}},\ \bibinfo {pages} {076501} (\bibinfo
  {year} {2012})}\BibitemShut {NoStop}%
\bibitem [{\citenamefont {Sato}\ and\ \citenamefont
  {Fujimoto}(2016)}]{Sato_JPSJ16}%
  \BibitemOpen
  \bibfield  {author} {\bibinfo {author} {\bibfnamefont {M.}~\bibnamefont
  {Sato}}\ and\ \bibinfo {author} {\bibfnamefont {S.}~\bibnamefont
  {Fujimoto}},\ }\href {\doibase 10.7566/JPSJ.85.072001} {\bibfield  {journal}
  {\bibinfo  {journal} {Journal of the Physical Society of Japan}\ }\textbf
  {\bibinfo {volume} {85}},\ \bibinfo {pages} {072001} (\bibinfo {year}
  {2016})},\ \Eprint
  {http://arxiv.org/abs/https://doi.org/10.7566/JPSJ.85.072001}
  {https://doi.org/10.7566/JPSJ.85.072001} \BibitemShut {NoStop}%
\bibitem [{\citenamefont {Haldane}\ and\ \citenamefont
  {Raghu}(2008)}]{Haldane_chiralPHC_PRL08}%
  \BibitemOpen
  \bibfield  {author} {\bibinfo {author} {\bibfnamefont {F.~D.~M.}\
  \bibnamefont {Haldane}}\ and\ \bibinfo {author} {\bibfnamefont
  {S.}~\bibnamefont {Raghu}},\ }\href {\doibase 10.1103/PhysRevLett.100.013904}
  {\bibfield  {journal} {\bibinfo  {journal} {Phys. Rev. Lett.}\ }\textbf
  {\bibinfo {volume} {100}},\ \bibinfo {pages} {013904} (\bibinfo {year}
  {2008})}\BibitemShut {NoStop}%
\bibitem [{\citenamefont {Raghu}\ and\ \citenamefont
  {Haldane}(2008)}]{Raghu_chiralPHC_PRA08}%
  \BibitemOpen
  \bibfield  {author} {\bibinfo {author} {\bibfnamefont {S.}~\bibnamefont
  {Raghu}}\ and\ \bibinfo {author} {\bibfnamefont {F.~D.~M.}\ \bibnamefont
  {Haldane}},\ }\href {\doibase 10.1103/PhysRevA.78.033834} {\bibfield
  {journal} {\bibinfo  {journal} {Phys. Rev. A}\ }\textbf {\bibinfo {volume}
  {78}},\ \bibinfo {pages} {033834} (\bibinfo {year} {2008})}\BibitemShut
  {NoStop}%
\bibitem [{\citenamefont {Wang}\ \emph {et~al.}(2009)\citenamefont {Wang},
  \citenamefont {Chong}, \citenamefont {Joannopoulos},\ and\ \citenamefont
  {Soljacic}}]{Wang_chiralPHC_Nature09}%
  \BibitemOpen
  \bibfield  {author} {\bibinfo {author} {\bibfnamefont {Z.}~\bibnamefont
  {Wang}}, \bibinfo {author} {\bibfnamefont {Y.}~\bibnamefont {Chong}},
  \bibinfo {author} {\bibfnamefont {J.~D.}\ \bibnamefont {Joannopoulos}}, \
  and\ \bibinfo {author} {\bibfnamefont {M.}~\bibnamefont {Soljacic}},\ }\href
  {https://doi.org/10.1038/nature08293} {\bibfield  {journal} {\bibinfo
  {journal} {Nature}\ }\textbf {\bibinfo {volume} {461}},\ \bibinfo {pages}
  {772 EP } (\bibinfo {year} {2009})}\BibitemShut {NoStop}%
\bibitem [{\citenamefont {Ozawa}\ \emph {et~al.}(2019)\citenamefont {Ozawa},
  \citenamefont {Price}, \citenamefont {Amo}, \citenamefont {Goldman},
  \citenamefont {Hafezi}, \citenamefont {Lu}, \citenamefont {Rechtsman},
  \citenamefont {Schuster}, \citenamefont {Simon}, \citenamefont {Zilberberg},\
  and\ \citenamefont {Carusotto}}]{Ozawa_TopoPhoto_RMP19}%
  \BibitemOpen
  \bibfield  {author} {\bibinfo {author} {\bibfnamefont {T.}~\bibnamefont
  {Ozawa}}, \bibinfo {author} {\bibfnamefont {H.~M.}\ \bibnamefont {Price}},
  \bibinfo {author} {\bibfnamefont {A.}~\bibnamefont {Amo}}, \bibinfo {author}
  {\bibfnamefont {N.}~\bibnamefont {Goldman}}, \bibinfo {author} {\bibfnamefont
  {M.}~\bibnamefont {Hafezi}}, \bibinfo {author} {\bibfnamefont
  {L.}~\bibnamefont {Lu}}, \bibinfo {author} {\bibfnamefont {M.~C.}\
  \bibnamefont {Rechtsman}}, \bibinfo {author} {\bibfnamefont {D.}~\bibnamefont
  {Schuster}}, \bibinfo {author} {\bibfnamefont {J.}~\bibnamefont {Simon}},
  \bibinfo {author} {\bibfnamefont {O.}~\bibnamefont {Zilberberg}}, \ and\
  \bibinfo {author} {\bibfnamefont {I.}~\bibnamefont {Carusotto}},\ }\href
  {\doibase 10.1103/RevModPhys.91.015006} {\bibfield  {journal} {\bibinfo
  {journal} {Rev. Mod. Phys.}\ }\textbf {\bibinfo {volume} {91}},\ \bibinfo
  {pages} {015006} (\bibinfo {year} {2019})}\BibitemShut {NoStop}%
\bibitem [{\citenamefont {Prodan}\ and\ \citenamefont
  {Prodan}(2009)}]{ProdanPRL09}%
  \BibitemOpen
  \bibfield  {author} {\bibinfo {author} {\bibfnamefont {E.}~\bibnamefont
  {Prodan}}\ and\ \bibinfo {author} {\bibfnamefont {C.}~\bibnamefont
  {Prodan}},\ }\href {\doibase 10.1103/PhysRevLett.103.248101} {\bibfield
  {journal} {\bibinfo  {journal} {Phys. Rev. Lett.}\ }\textbf {\bibinfo
  {volume} {103}},\ \bibinfo {pages} {248101} (\bibinfo {year}
  {2009})}\BibitemShut {NoStop}%
\bibitem [{\citenamefont {Kane}\ and\ \citenamefont
  {Lubensky}(2013)}]{Kane_NatPhys13}%
  \BibitemOpen
  \bibfield  {author} {\bibinfo {author} {\bibfnamefont {C.~L.}\ \bibnamefont
  {Kane}}\ and\ \bibinfo {author} {\bibfnamefont {T.~C.}\ \bibnamefont
  {Lubensky}},\ }\href {https://doi.org/10.1038/nphys2835} {\bibfield
  {journal} {\bibinfo  {journal} {Nature Physics}\ }\textbf {\bibinfo {volume}
  {10}},\ \bibinfo {pages} {39 EP } (\bibinfo {year} {2013})},\ \bibinfo {note}
  {article}\BibitemShut {NoStop}%
\bibitem [{\citenamefont {Kariyado}\ and\ \citenamefont
  {Hatsugai}(2015)}]{Kariyado_SR15}%
  \BibitemOpen
  \bibfield  {author} {\bibinfo {author} {\bibfnamefont {T.}~\bibnamefont
  {Kariyado}}\ and\ \bibinfo {author} {\bibfnamefont {Y.}~\bibnamefont
  {Hatsugai}},\ }\href {https://doi.org/10.1038/srep18107} {\bibfield
  {journal} {\bibinfo  {journal} {Scientific Reports}\ }\textbf {\bibinfo
  {volume} {5}},\ \bibinfo {pages} {18107 EP } (\bibinfo {year} {2015})},\
  \bibinfo {note} {article}\BibitemShut {NoStop}%
\bibitem [{\citenamefont {S{\"u}sstrunk}\ and\ \citenamefont
  {Huber}(2016)}]{Suesstrunk_Mech-class_PNAS16}%
  \BibitemOpen
  \bibfield  {author} {\bibinfo {author} {\bibfnamefont {R.}~\bibnamefont
  {S{\"u}sstrunk}}\ and\ \bibinfo {author} {\bibfnamefont {S.~D.}\ \bibnamefont
  {Huber}},\ }\href {\doibase 10.1073/pnas.1605462113} {\bibfield  {journal}
  {\bibinfo  {journal} {Proceedings of the National Academy of Sciences}\
  }\textbf {\bibinfo {volume} {113}},\ \bibinfo {pages} {E4767} (\bibinfo
  {year} {2016})}\BibitemShut {NoStop}%
\bibitem [{\citenamefont {Chien}\ \emph {et~al.}(2018)\citenamefont {Chien},
  \citenamefont {Velizhanin}, \citenamefont {Dubi}, \citenamefont {Ilic},\ and\
  \citenamefont {Zwolak}}]{Chien_Th_Ph_PRB18}%
  \BibitemOpen
  \bibfield  {author} {\bibinfo {author} {\bibfnamefont {C.-C.}\ \bibnamefont
  {Chien}}, \bibinfo {author} {\bibfnamefont {K.~A.}\ \bibnamefont
  {Velizhanin}}, \bibinfo {author} {\bibfnamefont {Y.}~\bibnamefont {Dubi}},
  \bibinfo {author} {\bibfnamefont {B.~R.}\ \bibnamefont {Ilic}}, \ and\
  \bibinfo {author} {\bibfnamefont {M.}~\bibnamefont {Zwolak}},\ }\href
  {\doibase 10.1103/PhysRevB.97.125425} {\bibfield  {journal} {\bibinfo
  {journal} {Phys. Rev. B}\ }\textbf {\bibinfo {volume} {97}},\ \bibinfo
  {pages} {125425} (\bibinfo {year} {2018})}\BibitemShut {NoStop}%
\bibitem [{\citenamefont {Yoshida}\ and\ \citenamefont
  {Hatsugai}(2019)}]{Yoshida_SPERs_mech19}%
  \BibitemOpen
  \bibfield  {author} {\bibinfo {author} {\bibfnamefont {T.}~\bibnamefont
  {Yoshida}}\ and\ \bibinfo {author} {\bibfnamefont {Y.}~\bibnamefont
  {Hatsugai}},\ }\href {\doibase 10.1103/PhysRevB.100.054109} {\bibfield
  {journal} {\bibinfo  {journal} {Phys. Rev. B}\ }\textbf {\bibinfo {volume}
  {100}},\ \bibinfo {pages} {054109} (\bibinfo {year} {2019})}\BibitemShut
  {NoStop}%
\bibitem [{\citenamefont {Wakao}\ \emph {et~al.}(2020)\citenamefont {Wakao},
  \citenamefont {Yoshida}, \citenamefont {Araki}, \citenamefont {Mizoguchi},\
  and\ \citenamefont {Hatsugai}}]{Wakao_HOTImech_PRB20}%
  \BibitemOpen
  \bibfield  {author} {\bibinfo {author} {\bibfnamefont {H.}~\bibnamefont
  {Wakao}}, \bibinfo {author} {\bibfnamefont {T.}~\bibnamefont {Yoshida}},
  \bibinfo {author} {\bibfnamefont {H.}~\bibnamefont {Araki}}, \bibinfo
  {author} {\bibfnamefont {T.}~\bibnamefont {Mizoguchi}}, \ and\ \bibinfo
  {author} {\bibfnamefont {Y.}~\bibnamefont {Hatsugai}},\ }\href {\doibase
  10.1103/PhysRevB.101.094107} {\bibfield  {journal} {\bibinfo  {journal}
  {Phys. Rev. B}\ }\textbf {\bibinfo {volume} {101}},\ \bibinfo {pages}
  {094107} (\bibinfo {year} {2020})}\BibitemShut {NoStop}%
\bibitem [{\citenamefont {Albert}\ \emph {et~al.}(2015)\citenamefont {Albert},
  \citenamefont {Glazman},\ and\ \citenamefont
  {Jiang}}]{Victor_Topoelecircit_PRL15}%
  \BibitemOpen
  \bibfield  {author} {\bibinfo {author} {\bibfnamefont {V.~V.}\ \bibnamefont
  {Albert}}, \bibinfo {author} {\bibfnamefont {L.~I.}\ \bibnamefont {Glazman}},
  \ and\ \bibinfo {author} {\bibfnamefont {L.}~\bibnamefont {Jiang}},\ }\href
  {\doibase 10.1103/PhysRevLett.114.173902} {\bibfield  {journal} {\bibinfo
  {journal} {Phys. Rev. Lett.}\ }\textbf {\bibinfo {volume} {114}},\ \bibinfo
  {pages} {173902} (\bibinfo {year} {2015})}\BibitemShut {NoStop}%
\bibitem [{\citenamefont {Lee}\ \emph {et~al.}(2018)\citenamefont {Lee},
  \citenamefont {Imhof}, \citenamefont {Berger}, \citenamefont {Bayer},
  \citenamefont {Brehm}, \citenamefont {Molenkamp}, \citenamefont {Kiessling},\
  and\ \citenamefont {Thomale}}]{Lee_Topoelecircit_CommPhys18}%
  \BibitemOpen
  \bibfield  {author} {\bibinfo {author} {\bibfnamefont {C.~H.}\ \bibnamefont
  {Lee}}, \bibinfo {author} {\bibfnamefont {S.}~\bibnamefont {Imhof}}, \bibinfo
  {author} {\bibfnamefont {C.}~\bibnamefont {Berger}}, \bibinfo {author}
  {\bibfnamefont {F.}~\bibnamefont {Bayer}}, \bibinfo {author} {\bibfnamefont
  {J.}~\bibnamefont {Brehm}}, \bibinfo {author} {\bibfnamefont {L.~W.}\
  \bibnamefont {Molenkamp}}, \bibinfo {author} {\bibfnamefont {T.}~\bibnamefont
  {Kiessling}}, \ and\ \bibinfo {author} {\bibfnamefont {R.}~\bibnamefont
  {Thomale}},\ }\href {\doibase 10.1038/s42005-018-0035-2} {\bibfield
  {journal} {\bibinfo  {journal} {Communications Physics}\ }\textbf {\bibinfo
  {volume} {1}},\ \bibinfo {pages} {39} (\bibinfo {year} {2018})}\BibitemShut
  {NoStop}%
\bibitem [{\citenamefont {Helbig}\ \emph {et~al.}(2019)\citenamefont {Helbig},
  \citenamefont {Hofmann}, \citenamefont {Imhof}, \citenamefont {Abdelghany},
  \citenamefont {Kiessling}, \citenamefont {Molenkamp}, \citenamefont {Lee},
  \citenamefont {Szameit}, \citenamefont {Greiter},\ and\ \citenamefont
  {Thomale}}]{Helbig_ExpSkin_19}%
  \BibitemOpen
  \bibfield  {author} {\bibinfo {author} {\bibfnamefont {T.}~\bibnamefont
  {Helbig}}, \bibinfo {author} {\bibfnamefont {T.}~\bibnamefont {Hofmann}},
  \bibinfo {author} {\bibfnamefont {S.}~\bibnamefont {Imhof}}, \bibinfo
  {author} {\bibfnamefont {M.}~\bibnamefont {Abdelghany}}, \bibinfo {author}
  {\bibfnamefont {T.}~\bibnamefont {Kiessling}}, \bibinfo {author}
  {\bibfnamefont {L.~W.}\ \bibnamefont {Molenkamp}}, \bibinfo {author}
  {\bibfnamefont {C.~H.}\ \bibnamefont {Lee}}, \bibinfo {author} {\bibfnamefont
  {A.}~\bibnamefont {Szameit}}, \bibinfo {author} {\bibfnamefont
  {M.}~\bibnamefont {Greiter}}, \ and\ \bibinfo {author} {\bibfnamefont
  {R.}~\bibnamefont {Thomale}},\ }\href@noop {} {\bibfield  {journal} {\bibinfo
   {journal} {arXiv preprint arXiv:1907.11562}\ } (\bibinfo {year}
  {2019})}\BibitemShut {NoStop}%
\bibitem [{\citenamefont {Yoshida}\ \emph {et~al.}(2020)\citenamefont
  {Yoshida}, \citenamefont {Mizoguchi},\ and\ \citenamefont
  {Hatsugai}}]{Yoshida_MSkinPRR20}%
  \BibitemOpen
  \bibfield  {author} {\bibinfo {author} {\bibfnamefont {T.}~\bibnamefont
  {Yoshida}}, \bibinfo {author} {\bibfnamefont {T.}~\bibnamefont {Mizoguchi}},
  \ and\ \bibinfo {author} {\bibfnamefont {Y.}~\bibnamefont {Hatsugai}},\
  }\href {\doibase 10.1103/PhysRevResearch.2.022062} {\bibfield  {journal}
  {\bibinfo  {journal} {Phys. Rev. Research}\ }\textbf {\bibinfo {volume}
  {2}},\ \bibinfo {pages} {022062} (\bibinfo {year} {2020})}\BibitemShut
  {NoStop}%
\bibitem [{\citenamefont {Delplace}\ \emph {et~al.}(2017)\citenamefont
  {Delplace}, \citenamefont {Marston},\ and\ \citenamefont
  {Venaille}}]{Delplace_topoEq_Science17}%
  \BibitemOpen
  \bibfield  {author} {\bibinfo {author} {\bibfnamefont {P.}~\bibnamefont
  {Delplace}}, \bibinfo {author} {\bibfnamefont {J.~B.}\ \bibnamefont
  {Marston}}, \ and\ \bibinfo {author} {\bibfnamefont {A.}~\bibnamefont
  {Venaille}},\ }\href {\doibase 10.1126/science.aan8819} {\bibfield  {journal}
  {\bibinfo  {journal} {Science}\ }\textbf {\bibinfo {volume} {358}},\ \bibinfo
  {pages} {1075} (\bibinfo {year} {2017})}\BibitemShut {NoStop}%
\bibitem [{\citenamefont {Sone}\ and\ \citenamefont
  {Ashida}(2019)}]{ActiveMatter_SonePRL19}%
  \BibitemOpen
  \bibfield  {author} {\bibinfo {author} {\bibfnamefont {K.}~\bibnamefont
  {Sone}}\ and\ \bibinfo {author} {\bibfnamefont {Y.}~\bibnamefont {Ashida}},\
  }\href {\doibase 10.1103/PhysRevLett.123.205502} {\bibfield  {journal}
  {\bibinfo  {journal} {Phys. Rev. Lett.}\ }\textbf {\bibinfo {volume} {123}},\
  \bibinfo {pages} {205502} (\bibinfo {year} {2019})}\BibitemShut {NoStop}%
\bibitem [{\citenamefont {Harari}\ \emph {et~al.}(2018)\citenamefont {Harari},
  \citenamefont {Bandres}, \citenamefont {Lumer}, \citenamefont {Rechtsman},
  \citenamefont {Chong}, \citenamefont {Khajavikhan}, \citenamefont
  {Christodoulides},\ and\ \citenamefont {Segev}}]{Harari_TopoLaser_Science18}%
  \BibitemOpen
  \bibfield  {author} {\bibinfo {author} {\bibfnamefont {G.}~\bibnamefont
  {Harari}}, \bibinfo {author} {\bibfnamefont {M.~A.}\ \bibnamefont {Bandres}},
  \bibinfo {author} {\bibfnamefont {Y.}~\bibnamefont {Lumer}}, \bibinfo
  {author} {\bibfnamefont {M.~C.}\ \bibnamefont {Rechtsman}}, \bibinfo {author}
  {\bibfnamefont {Y.~D.}\ \bibnamefont {Chong}}, \bibinfo {author}
  {\bibfnamefont {M.}~\bibnamefont {Khajavikhan}}, \bibinfo {author}
  {\bibfnamefont {D.~N.}\ \bibnamefont {Christodoulides}}, \ and\ \bibinfo
  {author} {\bibfnamefont {M.}~\bibnamefont {Segev}},\ }\href {\doibase
  10.1126/science.aar4003} {\bibfield  {journal} {\bibinfo  {journal}
  {Science}\ }\textbf {\bibinfo {volume} {359}} (\bibinfo {year} {2018}),\
  10.1126/science.aar4003}\BibitemShut {NoStop}%
\bibitem [{\citenamefont {Bandres}\ \emph {et~al.}(2018)\citenamefont
  {Bandres}, \citenamefont {Wittek}, \citenamefont {Harari}, \citenamefont
  {Parto}, \citenamefont {Ren}, \citenamefont {Segev}, \citenamefont
  {Christodoulides},\ and\ \citenamefont
  {Khajavikhan}}]{Banders_TopoLaser_Science18}%
  \BibitemOpen
  \bibfield  {author} {\bibinfo {author} {\bibfnamefont {M.~A.}\ \bibnamefont
  {Bandres}}, \bibinfo {author} {\bibfnamefont {S.}~\bibnamefont {Wittek}},
  \bibinfo {author} {\bibfnamefont {G.}~\bibnamefont {Harari}}, \bibinfo
  {author} {\bibfnamefont {M.}~\bibnamefont {Parto}}, \bibinfo {author}
  {\bibfnamefont {J.}~\bibnamefont {Ren}}, \bibinfo {author} {\bibfnamefont
  {M.}~\bibnamefont {Segev}}, \bibinfo {author} {\bibfnamefont {D.~N.}\
  \bibnamefont {Christodoulides}}, \ and\ \bibinfo {author} {\bibfnamefont
  {M.}~\bibnamefont {Khajavikhan}},\ }\href {\doibase 10.1126/science.aar4005}
  {\bibfield  {journal} {\bibinfo  {journal} {Science}\ }\textbf {\bibinfo
  {volume} {359}} (\bibinfo {year} {2018}),\
  10.1126/science.aar4005}\BibitemShut {NoStop}%
\bibitem [{\citenamefont {Ogi}\ \emph {et~al.}(2016)\citenamefont {Ogi},
  \citenamefont {Ishihara}, \citenamefont {Ishida}, \citenamefont {Nagakubo},
  \citenamefont {Nakamura},\ and\ \citenamefont {Hirao}}]{Ogi_DiffWavePRL16}%
  \BibitemOpen
  \bibfield  {author} {\bibinfo {author} {\bibfnamefont {H.}~\bibnamefont
  {Ogi}}, \bibinfo {author} {\bibfnamefont {T.}~\bibnamefont {Ishihara}},
  \bibinfo {author} {\bibfnamefont {H.}~\bibnamefont {Ishida}}, \bibinfo
  {author} {\bibfnamefont {A.}~\bibnamefont {Nagakubo}}, \bibinfo {author}
  {\bibfnamefont {N.}~\bibnamefont {Nakamura}}, \ and\ \bibinfo {author}
  {\bibfnamefont {M.}~\bibnamefont {Hirao}},\ }\href {\doibase
  10.1103/PhysRevLett.117.195901} {\bibfield  {journal} {\bibinfo  {journal}
  {Phys. Rev. Lett.}\ }\textbf {\bibinfo {volume} {117}},\ \bibinfo {pages}
  {195901} (\bibinfo {year} {2016})}\BibitemShut {NoStop}%
\bibitem [{\citenamefont {Li}\ \emph {et~al.}(2019)\citenamefont {Li},
  \citenamefont {Peng}, \citenamefont {Han}, \citenamefont {Miri},
  \citenamefont {Li}, \citenamefont {Xiao}, \citenamefont {Zhu}, \citenamefont
  {Zhao}, \citenamefont {Al{\`u}}, \citenamefont {Fan},\ and\ \citenamefont
  {Qiu}}]{LiSciencePT19}%
  \BibitemOpen
  \bibfield  {author} {\bibinfo {author} {\bibfnamefont {Y.}~\bibnamefont
  {Li}}, \bibinfo {author} {\bibfnamefont {Y.-G.}\ \bibnamefont {Peng}},
  \bibinfo {author} {\bibfnamefont {L.}~\bibnamefont {Han}}, \bibinfo {author}
  {\bibfnamefont {M.-A.}\ \bibnamefont {Miri}}, \bibinfo {author}
  {\bibfnamefont {W.}~\bibnamefont {Li}}, \bibinfo {author} {\bibfnamefont
  {M.}~\bibnamefont {Xiao}}, \bibinfo {author} {\bibfnamefont {X.-F.}\
  \bibnamefont {Zhu}}, \bibinfo {author} {\bibfnamefont {J.}~\bibnamefont
  {Zhao}}, \bibinfo {author} {\bibfnamefont {A.}~\bibnamefont {Al{\`u}}},
  \bibinfo {author} {\bibfnamefont {S.}~\bibnamefont {Fan}}, \ and\ \bibinfo
  {author} {\bibfnamefont {C.-W.}\ \bibnamefont {Qiu}},\ }\href {\doibase
  10.1126/science.aaw6259} {\bibfield  {journal} {\bibinfo  {journal}
  {Science}\ }\textbf {\bibinfo {volume} {364}},\ \bibinfo {pages} {170}
  (\bibinfo {year} {2019})},\ \Eprint
  {http://arxiv.org/abs/https://science.sciencemag.org/content/364/6436/170.full.pdf}
  {https://science.sciencemag.org/content/364/6436/170.full.pdf} \BibitemShut
  {NoStop}%
\bibitem [{\citenamefont {Peterson}\ and\ \citenamefont
  {Rothman}(1970)}]{Peterson_DiffImpMetal_PRB1970}%
  \BibitemOpen
  \bibfield  {author} {\bibinfo {author} {\bibfnamefont {N.~L.}\ \bibnamefont
  {Peterson}}\ and\ \bibinfo {author} {\bibfnamefont {S.~J.}\ \bibnamefont
  {Rothman}},\ }\href {\doibase 10.1103/PhysRevB.1.3264} {\bibfield  {journal}
  {\bibinfo  {journal} {Phys. Rev. B}\ }\textbf {\bibinfo {volume} {1}},\
  \bibinfo {pages} {3264} (\bibinfo {year} {1970})}\BibitemShut {NoStop}%
\bibitem [{\citenamefont {Su}\ \emph {et~al.}(1979)\citenamefont {Su},
  \citenamefont {Schrieffer},\ and\ \citenamefont {Heeger}}]{SSH_PRL79}%
  \BibitemOpen
  \bibfield  {author} {\bibinfo {author} {\bibfnamefont {W.~P.}\ \bibnamefont
  {Su}}, \bibinfo {author} {\bibfnamefont {J.~R.}\ \bibnamefont {Schrieffer}},
  \ and\ \bibinfo {author} {\bibfnamefont {A.~J.}\ \bibnamefont {Heeger}},\
  }\href {\doibase 10.1103/PhysRevLett.42.1698} {\bibfield  {journal} {\bibinfo
   {journal} {Phys. Rev. Lett.}\ }\textbf {\bibinfo {volume} {42}},\ \bibinfo
  {pages} {1698} (\bibinfo {year} {1979})}\BibitemShut {NoStop}%
\bibitem [{\citenamefont {Heeger}\ \emph {et~al.}(1988)\citenamefont {Heeger},
  \citenamefont {Kivelson}, \citenamefont {Schrieffer},\ and\ \citenamefont
  {Su}}]{SSH_RMP88}%
  \BibitemOpen
  \bibfield  {author} {\bibinfo {author} {\bibfnamefont {A.~J.}\ \bibnamefont
  {Heeger}}, \bibinfo {author} {\bibfnamefont {S.}~\bibnamefont {Kivelson}},
  \bibinfo {author} {\bibfnamefont {J.~R.}\ \bibnamefont {Schrieffer}}, \ and\
  \bibinfo {author} {\bibfnamefont {W.~P.}\ \bibnamefont {Su}},\ }\href
  {\doibase 10.1103/RevModPhys.60.781} {\bibfield  {journal} {\bibinfo
  {journal} {Rev. Mod. Phys.}\ }\textbf {\bibinfo {volume} {60}},\ \bibinfo
  {pages} {781} (\bibinfo {year} {1988})}\BibitemShut {NoStop}%
\bibitem [{sup()}]{supple}%
  \BibitemOpen
  \href@noop {} {}\bibinfo {note} {{ Supplemental material for details of the
  one-dimensional system with dimerization and the honeycomb lattice systems
  }}\BibitemShut {NoStop}%
\bibitem [{est()}]{estimate_ftnt}%
  \BibitemOpen
  \href@noop {} {}\bibinfo {note} {{ When the system is composed of aluminum,
  the half-life $\tau$ is estimated to be $ \tau \sim 1 \mathrm{ms}$. This can
  be seen as follows. Because our discretized equation should reproduce the
  thermal condition equation in the continuum limit, we can estimate the
  coefficient $D$ for $D=D'$. Namely, the thermal diffusivity of aluminum is
  approximately $\lambda_{\mathrm{Al}} \sim 1\times 10^{-4}
  \mathrm{m}^2/\mathrm{s}$ which can be estimated from data shown in
  Ref.~\onlinecite{Ogi_DiffWavePRL16}. Suppose that we can reproduce the
  continuum results with 10 sites for system whose length is $2 \mathrm{mm}$,
  $D$ should satisfy $Da^2=\lambda_{\mathrm{Al}}$ with $a$ denoting the
  distance of neighboring sites (i.e., $a=0.2\mathrm{mm}$ in this case).
  Therefore, $D$ is approximately, $D=2.5\times 10^{3} \mathrm{s}^{-1}$, which
  results in $ \tau \sim 1 \mathrm{ms}$ with $\tau \sim 1/D$ }}\BibitemShut
  {NoStop}%
\bibitem [{T_i()}]{T_is_au_ftnt}%
  \BibitemOpen
  \href@noop {} {}\bibinfo {note} {{ Concerning the temperature, specific
  choice of the unit does not matter because the ratio of the temperature is
  discussed throughout this paper }}\BibitemShut {NoStop}%
\bibitem [{\citenamefont {Fujita}\ \emph {et~al.}(1996)\citenamefont {Fujita},
  \citenamefont {Wakabayashi}, \citenamefont {Nakada},\ and\ \citenamefont
  {Kusakabe}}]{Fujita_FujitaState_JPSJ96}%
  \BibitemOpen
  \bibfield  {author} {\bibinfo {author} {\bibfnamefont {M.}~\bibnamefont
  {Fujita}}, \bibinfo {author} {\bibfnamefont {K.}~\bibnamefont {Wakabayashi}},
  \bibinfo {author} {\bibfnamefont {K.}~\bibnamefont {Nakada}}, \ and\ \bibinfo
  {author} {\bibfnamefont {K.}~\bibnamefont {Kusakabe}},\ }\href {\doibase
  10.1143/JPSJ.65.1920} {\bibfield  {journal} {\bibinfo  {journal} {Journal of
  the Physical Society of Japan}\ }\textbf {\bibinfo {volume} {65}},\ \bibinfo
  {pages} {1920} (\bibinfo {year} {1996})},\ \Eprint
  {http://arxiv.org/abs/https://doi.org/10.1143/JPSJ.65.1920}
  {https://doi.org/10.1143/JPSJ.65.1920} \BibitemShut {NoStop}%
\end{thebibliography}
%



\clearpage

\renewcommand{\thesection}{S\arabic{section}}
\renewcommand{\theequation}{S\arabic{equation}}
\setcounter{equation}{0}
\renewcommand{\thefigure}{S\arabic{figure}}
\setcounter{figure}{0}
\renewcommand{\thetable}{S\arabic{table}}
\setcounter{table}{0}
\makeatletter
\c@secnumdepth = 2
\makeatother

\onecolumngrid
\begin{center}
 {\large \textmd{Supplemental Materials:} \\[0.3em]
 {\bfseries 
 Bulk-edge correspondence of classical diffusion phenomena
 }
 }
\end{center}

\setcounter{page}{1}

\section{
Details of the SSH model
}
\label{sec: SSH app}

Here, we derive the heat conduction equation, Eq.~(\ref{eq: diff SSH}), for the system illustrated in Fig.~\ref{fig: SSH_spec}(a).

Firstly, let us start with the case of $D'=0$. 
In this case, the isolated site is coupled to the wall.
By making use of Fourier's law, the heat flux from site $(i_x,\alpha)=(0,A)$ to the wall is written as
\begin{eqnarray}
\vec{Q}_{0A\to \mathrm{w}}&=& -D (T_{0A}-T_{\mathrm{w}}),
\end{eqnarray}
where $D$ denotes the diffusion coefficient.
The temperatures at site $(i_x,\alpha)=(0,A)$ and the wall are denoted by $T_{0A}$ and $T_{\mathrm{w}}$, respectively.

Because the heat at each site is rewritten as the temperature with the heat capacity, we have
\begin{eqnarray}
\left(
\begin{array}{cc}
C_w & 0 \\
0 & C
\end{array}
\right)
\partial_t
\left(
\begin{array}{c}
T_{\mathrm{w}}  \\
T_{0A} 
\end{array}
\right)
&=& 
-D
\left(
\begin{array}{cc}
1 & -1 \\
-1 & 1
\end{array}
\right)
\left(
\begin{array}{c}
T_{\mathrm{w}}  \\
T_{0A} 
\end{array}
\right),
\nonumber \\
\end{eqnarray}
where $C_{\mathrm{w}}$ and $C$ denote the heat capacity of the wall and site $(i_x,\alpha)=(0,A)$.

By multiplying the matrix
$
\left(
\begin{array}{cc}
C_{\mathrm{w}} & 0 \\
0 & C
\end{array}
\right)^{-1}
$ 
from left, 
the above equation is rewritten as 
\begin{eqnarray}
\partial_t
\left(
\begin{array}{c}
T_{\mathrm{w}}  \\
T_{0A} 
\end{array}
\right)
&=& 
-D
\left(
\begin{array}{cc}
1/C_w & -1/C_w \\
-1/C & 1/C
\end{array}
\right)
\left(
\begin{array}{c}
T_{\mathrm{w}}  \\
T_{0A} 
\end{array}
\right).
\end{eqnarray}
When $C_{\mathrm{w}}$ is infinitely large, $T_{\mathrm{w}}$ becomes independent of time. 
With this approximation and defining $T_{\mathrm{w}}=0$, we have
\begin{eqnarray}
\partial_t
\left(
\begin{array}{c}
0  \\
T_{0A} 
\end{array}
\right)
&=& 
-\frac{D}{C}
\left(
\begin{array}{cc}
0  & 0 \\
-1 & 1
\end{array}
\right)
\left(
\begin{array}{c}
0  \\
T_{0A} 
\end{array}
\right).
\end{eqnarray}

In a similar way, we have the heat conduction equation Eq.~(\ref{eq: diff SSH}) for $D'\neq 0$.
Namely, the time-evolution of the temperatures
\begin{eqnarray}
\vec{T}&=&
\left(
\begin{array}{cccccc}
T_{0A} & T_{0B} & T_{1A} & \cdots & T_{L_x-1A} & T_{L_x-1B}
\end{array}
\right),
\end{eqnarray}
are given by
\begin{subequations}
\label{eq: diff SSH app}
\begin{eqnarray}
\partial_t \vec{T}(t) &=& -\hat{H}_{\mathrm{SSH}}\vec{T}(t),
\end{eqnarray}
%
%
%
\begin{eqnarray}
\hat{H}_{\mathrm{SSH}} &=& D
\left(
\begin{array}{cccccc}
1+\delta   & -1        &  0        &    \cdots & -\delta  \\
-1         & 1+\delta  & -\delta   &    \cdots & 0 \\
 0         & -\delta   & 1+\delta  &    \cdots & 0 \\
\vdots     &  \vdots   &  \vdots   & \ddots    &  \vdots \\
-\delta    &  0        &   0       &  \cdots   &  1+\delta \\
\end{array}
\right),
\nonumber \\
\end{eqnarray}
%
\end{subequations}
with $\delta:=D'/D$.
Here, we note that $\hat{H}_{\mathrm{SSH}}-D(1+\delta)\1$ is identical to the Su-Schrieffer-Heeger (SSH) model, the one-dimensional tight-binding model with the dimerization $\delta$.

\section{
Details of the honeycomb lattice model
}
\label{sec: honey app}

\subsection{
Spectrum of the honeycomb lattice model
}
\label{sec: honey spec app}

\begin{figure}[!h]
\begin{minipage}{0.49\hsize}
\begin{center}
\includegraphics[width=0.7\hsize,clip]{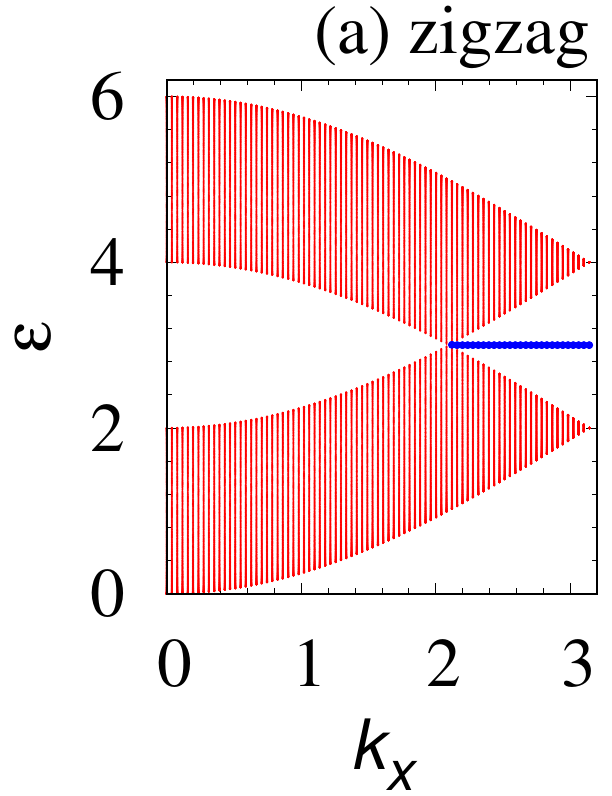}
\end{center}
\end{minipage}
\begin{minipage}{0.49\hsize}
\begin{center}
\includegraphics[width=0.7\hsize,clip]{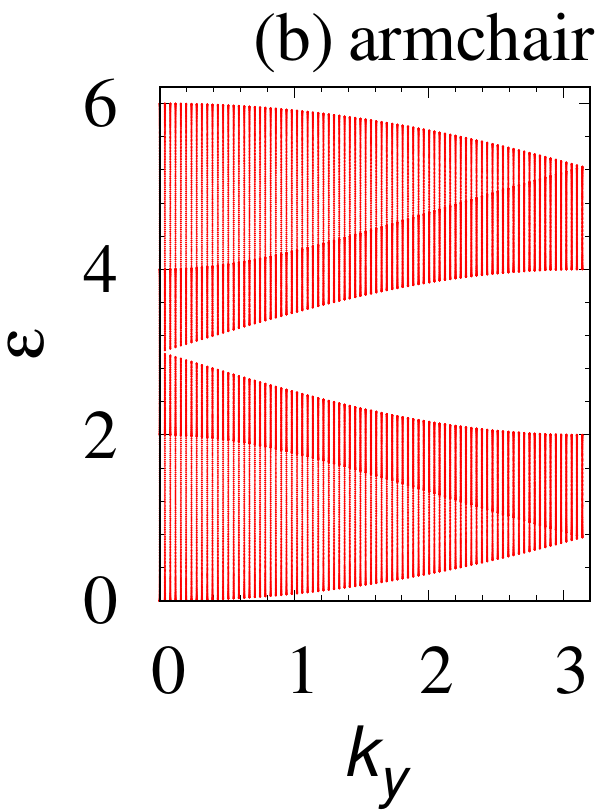}
\end{center}
\end{minipage}
\caption{(Color Online).
(a)[(b)]: The eigenvalues for the honeycomb lattice system with zigzag (armchair) edges.
For system with zigzag edges, we can find the edge states denoted by blue dots for $2\pi/3< k_x <\pi$.
We note that the spectrum is symmetric about $k_{x(y)}=0$.
The spectrum for zigzag edges (armchair edges) are obtained by imposing the periodic and the fixed (fixed and the periodic) boundary conditions for the $x$- and $y$-directions, respectively.
These data are obtained for $D=1$.
We suppose that $240$ unit cells are aligned along the direction where the fixed boundary condition is imposed.
}
\label{fig: honey_specOBC}
\end{figure}

The spectrum of honeycomb lattice model is plotted in Fig.~\ref{fig: honey_specOBC}. 
Figure~\ref{fig: honey_specOBC}(a) shows the spectrum under the periodic and fixed boundary conditions along the $x$- and $y$-directions.
In this case, applying the Fourier transformation along the $x$-direction, we can map the two-dimensional system to the one-dimensional system $H_{\mathrm{honey}}(k_x)$ parameterized by $k_x$.
As is the case of the SSH model, $H_{\mathrm{honey}}(k_x)$ preserves the chiral symmetry up to the term proportional to the identity matrix, which allows us to compute the winding number for each value of $k_x$.

In the case of the zigzag edge, the winding number takes $W=1$ for $ 2\pi/3 < k_x < \pi $, inducing the edge modes at $\epsilon = 3D $ [see Fig.~\ref{fig: honey_specOBC}(a)].

In the case of the zigzag edge, the winding number is always zero, and thus, no edge state is observed at $\epsilon =3D $ [see Fig.~\ref{fig: honey_specOBC}(b)].

\subsection{
Initial conditions
}
\label{sec: honey ini app}

\begin{figure}[!h]
\begin{minipage}{0.49\hsize}
\begin{center}
\includegraphics[width=0.7\hsize,clip]{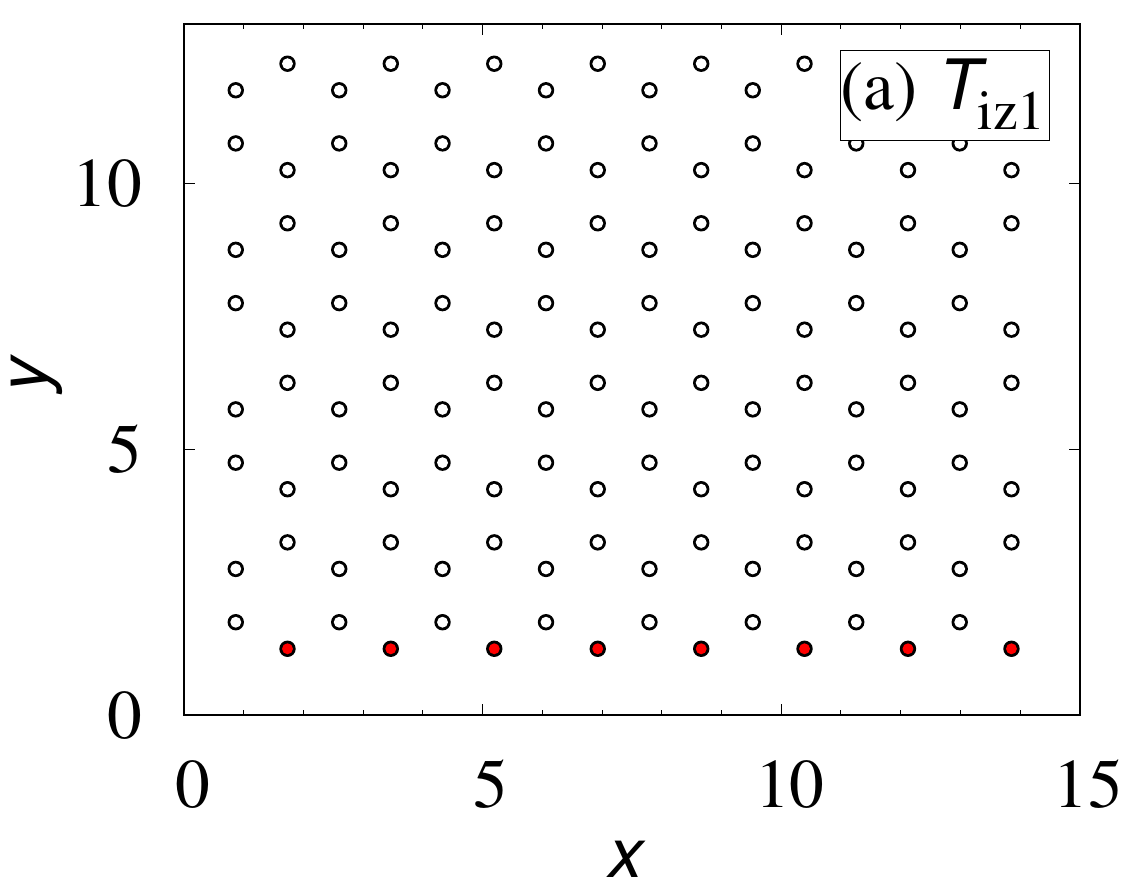}
\end{center}
\end{minipage}
\begin{minipage}{0.49\hsize}
\begin{center}
\includegraphics[width=0.7\hsize,clip]{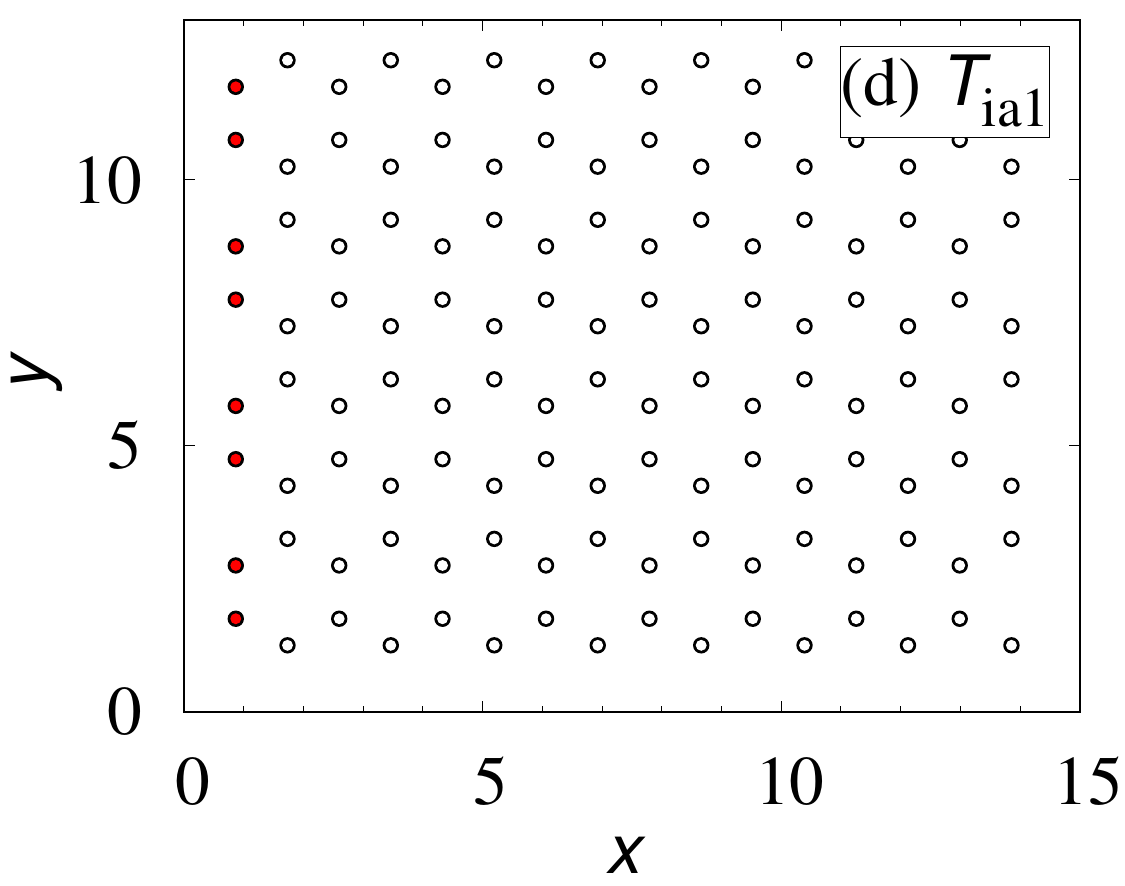}
\end{center}
\end{minipage}
\begin{minipage}{0.49\hsize}
\begin{center}
\includegraphics[width=0.7\hsize,clip]{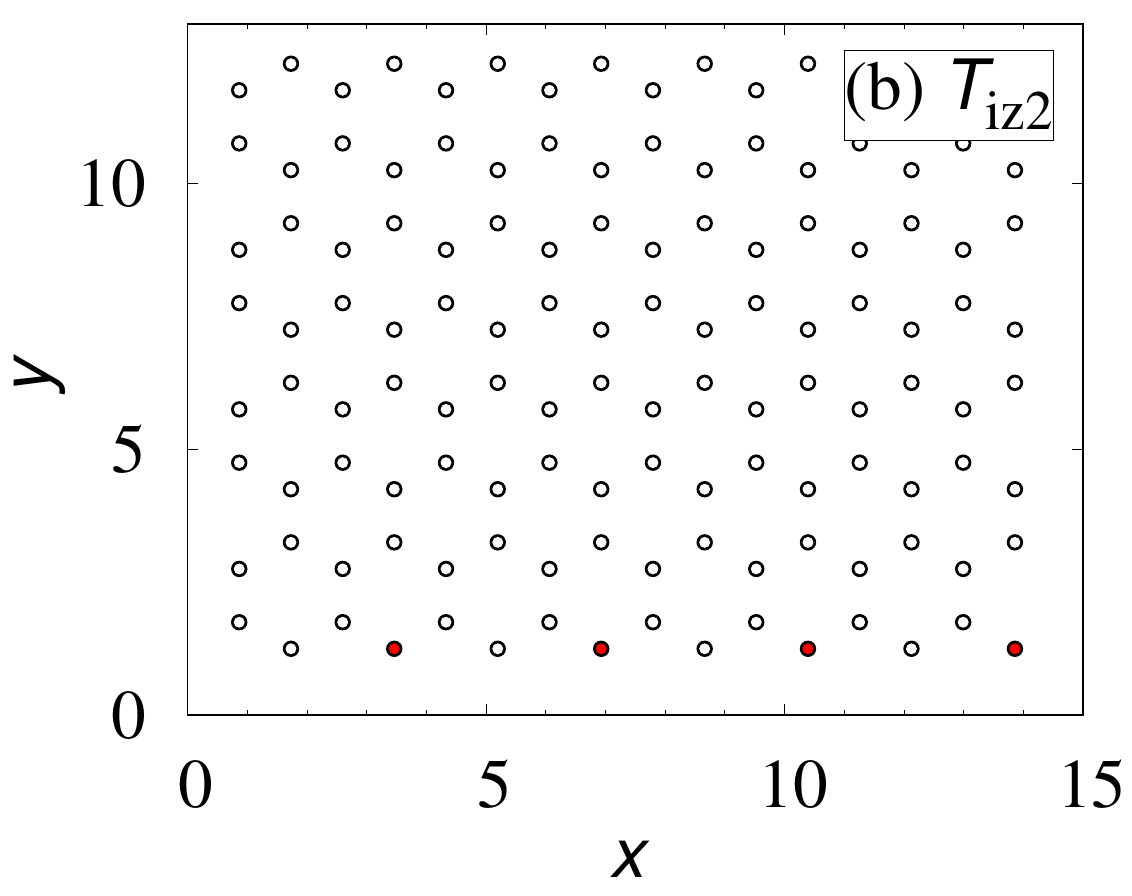}
\end{center}
\end{minipage}
\begin{minipage}{0.49\hsize}
\begin{center}
\includegraphics[width=0.7\hsize,clip]{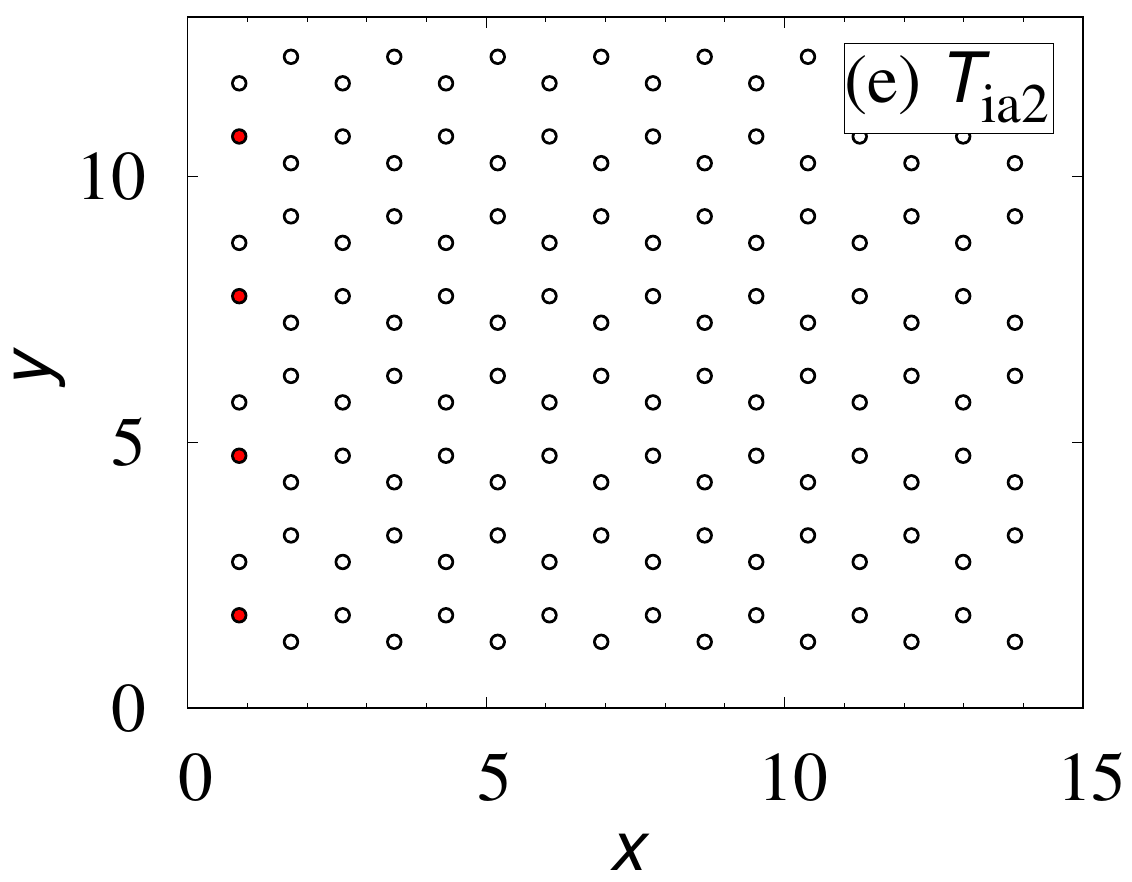}
\end{center}
\end{minipage}
\begin{minipage}{0.49\hsize}
\begin{center}
\includegraphics[width=0.7\hsize,clip]{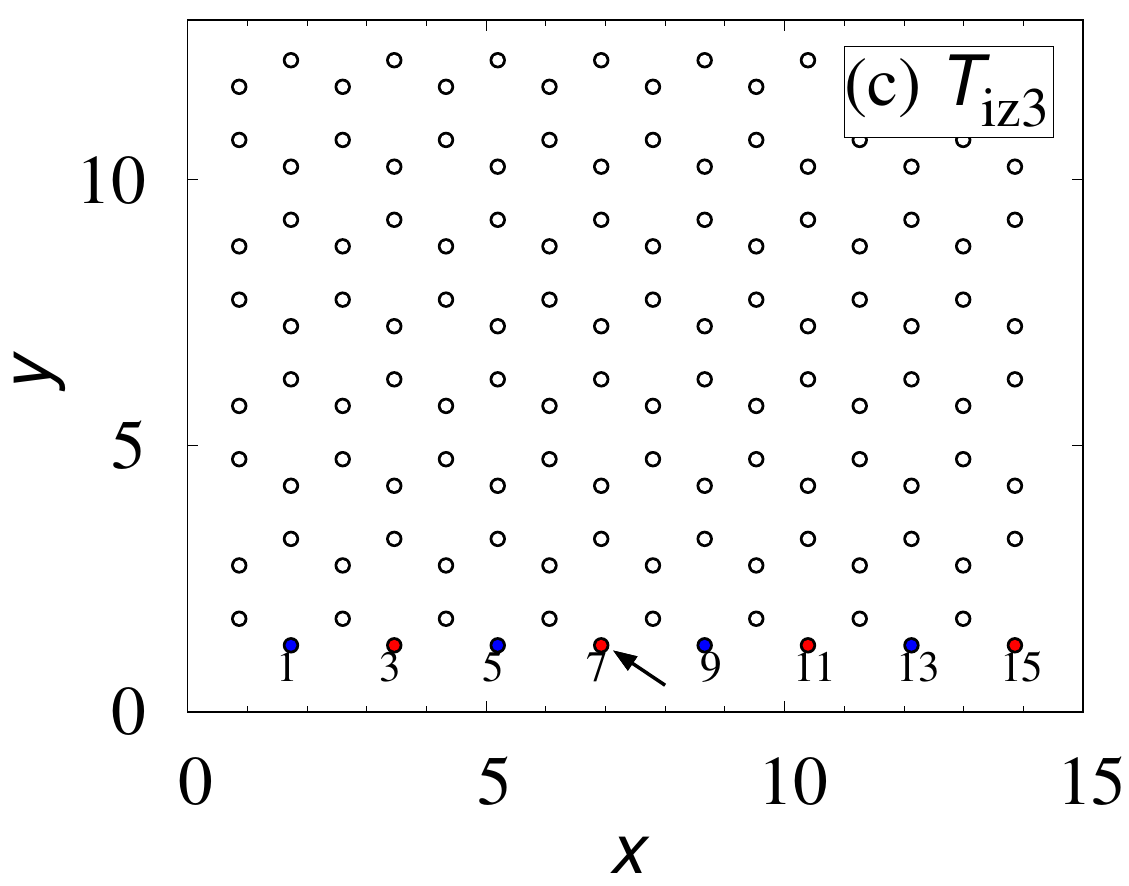}
\end{center}
\end{minipage}
\begin{minipage}{0.49\hsize}
\begin{center}
\includegraphics[width=0.7\hsize,clip]{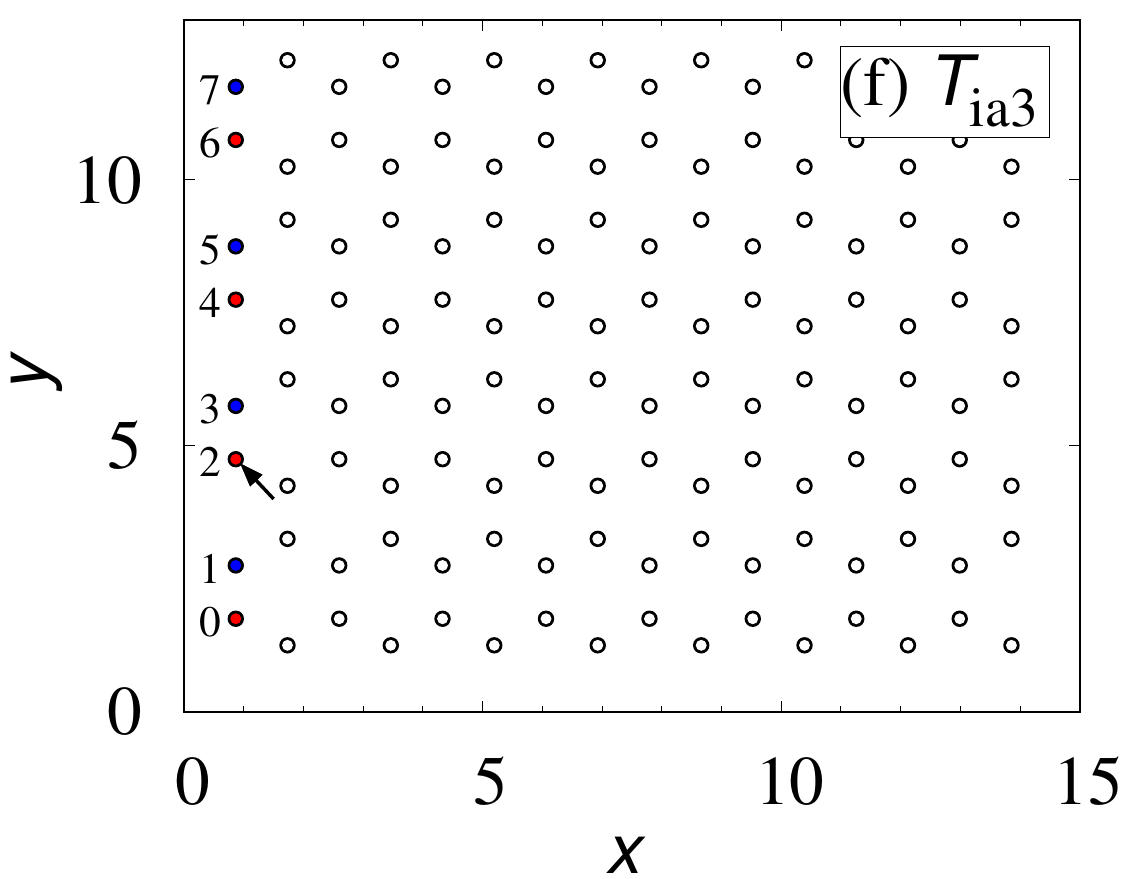}
\end{center}
\end{minipage}
\caption{(Color Online).
Color plot of vectors $\vec{T}$'s describing initial conditions for $L_x=16$ and $L_y=8$.
The red, white, and blue dots denote the points where the temperature is $1$, $0$, and $-1$, respectively.
In panel (c) [(f)], the site specified by $i_{_{\mathrm{cz}}}=(L_x/2-1,0)$ [$i_{\mathrm{ca}}=(0,L_y/2-2)$] is denoted by the arrow.
The sites labeled by $i_x=0,1,\cdots,L_x-1$ ($i_x=0,1,\cdots,L_y-1$) as shown in panel (c) [(f)].
}
\label{fig: honey_ini}
\end{figure}
In this section we explain the details of the initial conditions which we chose to obtain Figs.~\ref{fig: honeycomb}(b),~\ref{fig: honeycomb}(c),~and~\ref{fig: honeycomb_cmap}.

Figure~\ref{fig: honey_ini} shows vectors defining the initial conditions.
The data shown in Fig.~\ref{fig: honeycomb}(b) are obtained by simulating the dynamics for two cases of the initial condition: $\vec{T}_{\mathrm{iz1}}$ and $2\vec{T}_{\mathrm{iz2}}$. 
The data denoted by $k_x=0$ are obtained with the initial condition $\vec{T}_{\mathrm{iz1}}$. The data denoted by $k_x=\pi$ are obtained by subtracting the data with the initial condition $2\vec{T}_{\mathrm{iz2}}$ from the ones with $\vec{T}_{\mathrm{iz1}}$.
We note that the data labeled by $k_x=\pi$ are identical to the ones with the initial condition $\vec{T}_{\mathrm{iz3}}=\vec{T}_{\mathrm{iz1}}-2\vec{T}_{\mathrm{iz2}}$ because the diffusion equation is the linear equation.
The data shown in Fig.~\ref{fig: honeycomb}(c) are obtained by simulating the dynamics for two cases of the initial condition: $\vec{T}_{\mathrm{ia1}}$ or $2\vec{T}_{\mathrm{ia2}}$. Namely, the data denoted by $k_y=0$ ($k_y=0$) are obtained with the initial condition $\vec{T}_{\mathrm{ia1}}$ ($\vec{T}_{\mathrm{ia3}}=\vec{T}_{\mathrm{ia1}}-2\vec{T}_{\mathrm{ia2}}$).

Figures~\ref{fig: honeycomb_cmap}(a)~and~\ref{fig: honeycomb_cmap}(b) are obtained by simulating the time-evolution with the initial condition $\vec{T}_{\mathrm{iz3}}$ ($\vec{T}_{\mathrm{ia3}}$).

\end{document}